\documentclass[twoside]{article}

\usepackage[accepted]{aistats2024}
\usepackage{graphicx}
\usepackage{subcaption}
\usepackage{multirow}
\usepackage[utf8]{inputenc}
\usepackage{multicol}
\usepackage{hyperref}
\usepackage{amsmath}
\usepackage{amsfonts}

\begin{document}



\twocolumn[

\aistatstitle{BioDiffusion: A Versatile Diffusion Model for Biomedical Signal Synthesis}

\aistatsauthor{Xiaomin Li $^{1}$, Mykhailo Sakevych $^{1}$, Gentry Atkinson $^{2}$, and Vangelis Metsis $^{1}$ }

\aistatsaddress{Texas State University$^{1}$ xminli, ukb12, vmetsis@txstate.edu \\ St. Edwards University$^{2}$ gatkinso@stedwards.edu } ]

\begin{abstract}
Machine learning tasks involving biomedical signals frequently grapple with issues such as limited data availability, imbalanced datasets, labeling complexities, and the interference of measurement noise. These challenges often hinder the optimal training of machine learning algorithms. Addressing these concerns, we introduce BioDiffusion, a diffusion-based probabilistic model optimized for the synthesis of multivariate biomedical signals. BioDiffusion demonstrates excellence in producing high-fidelity, non-stationary, multivariate signals for a range of tasks including unconditional, label-conditional, and signal-conditional generation. Leveraging these synthesized signals offers a notable solution to the aforementioned challenges. Our research encompasses both qualitative and quantitative assessments of the synthesized data quality, underscoring its capacity to bolster accuracy in machine learning tasks tied to biomedical signals. Furthermore, when juxtaposed with current leading time-series generative models, empirical evidence suggests that BioDiffusion outperforms them in biomedical signal generation quality.

Source code link: \href{https://github.com/imics-lab/biodiffusion}{Link.}
\end{abstract}

\section{Introduction}
\label{sec:introduction}

The significance of biomedical signal processing is continually emphasized in its role across various ubiquitous computing applications. The quest for accurate, dependable data has been a driving force for innovations that lead to improved assistive technologies and deeper insights into diagnostics, patient monitoring, and therapeutics. Electrocardiograms (ECGs), electroencephalograms (EEGs), and data from human activity sensors represent a treasure trove of information. Their analysis has ushered in transformative breakthroughs, but not without associated challenges.

One major hurdle faced by biomedical signal processing is the intricacies that arise due to limited dataset size, imbalances in datasets, artificial noise, and anomalies. These factors can critically compromise the performance of machine learning models, necessitating the development of innovative solutions. Historically, approaches like data augmentation, data resampling, and statistical analyses have been employed to mitigate these challenges. Yet, the inherently non-stationary and multivariate characteristics of biomedical signals add another layer of complexity. Encouragingly, recent research trends highlight an uptick in leveraging deep learning for enhancing the preprocessing of biomedical signals~\cite{kachuee2018ecg, ronao2016human, roy2019deep}. 

Deep learning, though powerful, is often constrained by the nuances of biomedical datasets. Recognizing these challenges, our study introduces the BioDiffusion model, a diffusion-based probabilistic approach tailored for biomedical signal generation. Designed to adeptly handle a plethora of generation tasks, BioDiffusion serves as a holistic solution to biomedical signal synthesis challenges. From expanding training dataset sizes to anomaly removal and super-resolution, our model's adaptability offers a promising avenue for more efficient and precise analysis techniques in clinical applications.

Inspired by the Stable Diffusion model's prowess in image synthesis~\cite{rombach2021highresolution}, we've adapted the BioDiffusion model to resonate with the unique traits of biomedical signals. To evaluate our model, we've engaged in a multi-faceted assessment, employing visual similarity comparisons, dimension reduction technologies like UMAP~\cite{mcinnes2018umap}, and similarity scores such as wavelet coherence. Additionally, our research delves into the synthesis signals' potential in machine learning model training, juxtaposing synthetic signals against real signals.

Through rigorous benchmarking against contemporary time-series synthesis models, our findings illuminate the BioDiffusion model's superior proficiency in generating high-fidelity biomedical signals. The implications of our proposed model are profound; it presents a significant stride toward enhancing diagnostics, patient monitoring, and advancing biomedical research.

\noindent \textbf{Main Contributions:}
\begin{itemize}
    \item Presentation of the BioDiffusion model, our innovative diffusion-based probabilistic approach tailored to address the complexities inherent in biomedical signal generation.
    \item Demonstration of our model's versatility in handling diverse generation tasks, presenting a unified solution to biomedical signal synthesis.
    \item Comprehensive evaluation of the BioDiffusion model through both qualitative and quantitative metrics, underscoring its effectiveness and precision.
    \item Comparative analysis highlighting the superior capability of BioDiffusion in biomedical signal synthesis relative to existing state-of-the-art models.
\end{itemize}

The remainder of this paper is structured as follows: Section \ref{sec:related_work} delves into pertinent works related to signal synthesis. Section \ref{sec:methodology} elucidates the BioDiffusion model's methodology and its specific adaptions for biomedical signals. Section \ref{sec:experiments} presents our experimental framework, datasets, evaluation metrics, and a comparative analysis with other models, underlining BioDiffusion's standout performance. Finally, Section \ref{sec:conclusion} rounds off the paper, encapsulating the salient points and suggesting avenues for future exploration.

\section{Related Work}
\label{sec:related_work}

This section delves into the pertinent literature in the realms of generative models and diffusion models for signal synthesis. Our objective is to offer a comprehensive perspective on their evolution, strengths, and constraints, especially in the context of time-series signal synthesis.

\subsection{Generative Models in Signal Synthesis}

Generative models aim to discern the inherent structure of data, enabling the generation of new samples mirroring the original data. Several paradigmatic approaches within generative models for time-series synthesis include:

\begin{itemize}

\item \textbf{Generative Adversarial Networks (GANs)}: Comprising two adversarial networks—the generator and the discriminator—GANs aim for the generator to improve its synthetic data samples to deceive the discriminator. Their prowess extends to various data types including time-series signals. Notable implementations include the transformer-based GAN by Xiaomin L. et al.~\cite{li2022tts} which sets a benchmark for synthetic time-series signal fidelity, TimeGAN by Jinsung Y. et al.~\cite{yoon2019time} tailoring GANs for realistic time-series data, and the Recurrent Conditional GAN (RCGAN) by Cristóbal E. et al.~\cite{esteban2017real} for time-series generation. Despite their proficiency in crafting realistic samples, GANs can exhibit training instability and suffer from mode collapse.

\item \textbf{Variational Autoencoders (VAEs)}: VAEs, through their encoder-decoder architecture, learn a probabilistic representation of data. Works such as that by Vincent F. et al.~\cite{fortuin2019multivariate} exploit VAEs for imputing missing multivariate time-series values, while Fu et al.~\cite{dataaugmentation} leverage VAEs for augmenting time-series in human activity recognition. VAEs offer more consistent training than GANs but may produce less diverse samples, contingent on latent space distribution choices.

\item \textbf{Autoregressive Models}: These models sequentially generate samples, with each new element contingent on prior elements. WaveNet by Aaron van den Oord et al.~\cite{oord2016wavenet} exemplifies this, producing raw audio waveforms using dilated causal convolutions for long-range temporal relationship capture. Although proficient in modeling temporal dynamics, their sequential nature can be computationally slow and may falter in grasping extended dependencies.

\item Other generative paradigms like \textbf{Normalizing Flows}, \textbf{Restricted Boltzmann Machines (RBMs)}, and \textbf{Non-negative Matrix Factorization (NMF)} have been explored. However, their efficacy diminishes with multidimensional non-stationary time-series signals.

\end{itemize}

\subsection{Diffusion Models for Time-series Synthesis}

Diffusion models harness latent variables to understand a dataset by modeling data point propagation through latent space. They function by adding Gaussian noise to training data (forward diffusion) and subsequently reversing this process (reverse diffusion) to retrieve the data~\cite{yang2022diffusion}. Their prowess has been manifested in diverse arenas like image synthesis and molecule design~\cite{uniteai2021diffusion}. 

Several prominent studies in diffusion models include:
\begin{itemize}

\item Yang L. et al.'s comprehensive discourse on deep learning-based diffusion models and their applicability to time-series tasks~\cite{yang2022diffusion}.
  
\item Garnier O. et al. augmenting diffusion models for infinite-dimensional spaces, targeting audio signals and time series~\cite{garnier2023diffusion}.
  
\item Kong et al.'s exploration into audio synthesis through diffusion models~\cite{kong2020diffwave} and Tashiro et al.'s venture into time-series imputation~\cite{tashiro2021csdi}.
  
\item Alcaraz et al.'s pursuit of time-series forecasting using diffusion models~\cite{alcaraz2022diffusion}.

\end{itemize}

While these studies accentuate the capabilities of generative and diffusion models for time-series synthesis, there remain challenges in terms of scalability, stability, and fidelity, especially for intricate biomedical signals. Our proposed BioDiffusion model stands as an endeavor to surmount these challenges, deriving inspiration from prior works while innovating for enhanced versatility and efficacy in biomedical signal synthesis. The forthcoming section elucidates the methodology underlying BioDiffusion, illustrating its potential to revolutionize biomedical signal synthesis.

\section{Diffusion Probabilistic Models}
\label{sec:methodology}

This section provides an overview of diffusion models, their extension to conditional data generation, and associated neural architectures.

Diffusion models~\cite{sohl2015deep,ho2020denoising} consist of a forward process that iteratively degrades data $\mathbf{x}_0 \sim q(\mathbf{x}_0)$ by adding Gaussian noise over $T$ iterations:
\begin{align}
    q\left(\mathbf{x}_t \mid \mathbf{x}_{t-1}\right) &= \mathcal{N}\left(\mathbf{x}_t ; \sqrt{1-\beta_t} \mathbf{x}_{t-1}, \beta_t \mathbf{I}\right), \\
    q\left(\mathbf{x}_{1: T} \mid \mathbf{x}_0\right) &= \prod_{t=1}^T q\left(\mathbf{x}_t \mid \mathbf{x}_{t-1}\right).
\end{align}

The reverse process incrementally restores the noise-corrupted data:
\begin{align}
    p_\theta\left(\mathbf{x}_{t-1} \mid \mathbf{x}_t\right) &= \mathcal{N}\left(\mathbf{x}_{t-1} ; \boldsymbol{\mu}_\theta\left(\mathbf{x}_t, t\right), \boldsymbol{\Sigma}_\theta\left(\mathbf{x}_t, t\right)\right), \\
    p_\theta\left(\mathbf{x}_{0: T}\right) &= p\left(\mathbf{x}_T\right) \prod_{t=1}^T p_\theta\left(\mathbf{x}_{t-1} \mid \mathbf{x}_t\right).
\end{align}

The forward process hyperparameters $\beta_t$ are set such that $\mathbf{x}_T$ approximates a standard normal distribution. The reverse process optimizes the evidence lower bound (ELBO)~\cite{ganguly2021introduction}, with the loss given by:
\begin{equation}
\hspace{-4mm}
\resizebox{0.9\linewidth}{!}{%
  \begin{minipage}{\linewidth}
    \[
    \begin{aligned}
      L_\theta\left(\mathbf{x}_0\right) &= \mathbb{E}_q\left[L_T\left(\mathbf{x}_0\right) + \sum_{t>1} D_{\mathrm{KL}}\left(q\left(\mathbf{x}_{t-1} \mid \mathbf{x}_t, \mathbf{x}_0\right) \right. \right. \\
      &\left. \left. \| p_\theta\left(\mathbf{x}_{t-1} \mid \mathbf{x}_t\right)\right) - \log p_\theta\left(\mathbf{x}_0 \mid \mathbf{x}_1\right)\right],
    \end{aligned}
    \]
  \end{minipage}
}
\end{equation}

where $L_T\left(\mathbf{x}_0\right) = D_{\mathrm{KL}}\left(q\left(\mathbf{x}_T \mid \mathbf{x}_0\right) \| p\left(\mathbf{x}_T\right)\right)$.

Following prior work~\cite{sohl2015deep,ho2020denoising}, the reverse process parameters are:

\begin{align}
    \boldsymbol{\mu}_\theta\left(\mathbf{x}_t, t\right) &= \frac{1}{\sqrt{\alpha_t}}\left(\mathbf{x}_t-\frac{\beta_t}{\sqrt{1-\bar{\alpha}_t}} \boldsymbol{\epsilon}_\theta\left(\mathbf{x}_t, t\right)\right), \\
    \Sigma_\theta^{i i}\left(\mathbf{x}_t, t\right) &= \exp \left(\log \tilde{\beta}_t+\left(\log \beta_t-\log \tilde{\beta}_t\right) v_\theta^i\left(\mathbf{x}_t, t\right)\right),
\end{align}

with $\alpha_t=1-\beta_t$, $\bar{\alpha}_t=\prod_{s=1}^t \alpha_s$, and $\tilde{\beta}_t=\frac{1-\bar{\alpha}_{t-1}}{1-\bar{\alpha}_t} \beta_t$.

Improved sample quality is achieved by optimizing modified losses, resembling denoising score matching over multiple noise levels~\cite{song2020improved,ho2020denoising}.

A critical aspect of diffusion models is the extension to conditional data generation, wherein both the data \(\mathbf{x}_0\) and a set of conditions \(\mathbf{c}\) are incorporated. The conditions can be any additional information or constraints provided externally, influencing the generative process. By assimilating \(\mathbf{c}\), the reverse process becomes:

% \begin{align}
%     p_\theta\left(\mathbf{x}_{t-1} \mid \mathbf{x}_t, \mathbf{c}\right) &= \mathcal{N}\left(\mathbf{x}_{t-1} ; \boldsymbol{\mu}_\theta\left(\mathbf{x}_t, t, \mathbf{c}\right), \boldsymbol{\Sigma}_\theta\left(\mathbf{x}_t, t, \mathbf{c}\right)\right).
% \end{align}
\vspace{-5mm}
\begin{equation}
\hspace{-2mm}
\resizebox{0.9\linewidth}{!}{$ p_\theta\left(\mathbf{x}_{t-1} \mid \mathbf{x}_t, \mathbf{c}\right) \mathcal{N}\left(\mathbf{x}_{t-1} ; \boldsymbol{\mu}_\theta\left(\mathbf{x}_t, t, \mathbf{c}\right), \boldsymbol{\Sigma}_\theta\left(\mathbf{x}_t, t, \mathbf{c}\right)\right). $
}
\end{equation}

Intuitively, \(\mathbf{c}\) offers an avenue to guide the generative model, providing a degree of control over the outputs. This inclusion makes diffusion models versatile, catering to scenarios like content-specific image generation or style-conditioned audio synthesis.

For the architecture, we employ a feed-forward neural network. It has distinct input layers for data, the conditions \(\mathbf{c}\), and the time step. In line with the approach in [12], our model leverages multi-scale structures through convolutional layers, enabling the extraction of hierarchical information. The training strategy employs early stopping, hinging on the validation set ELBO to prevent overfitting.

\subsection{Unconditional Diffusion Models}
% In the unconditional diffusion process, the diffusion model treats the generation process as a Markov chain of steps that transform data into noise and vice versa. Figure~\ref{fig:unconditional_diffusion} illustrates how this process operates on signals. In the forward process, an original signal is progressively mixed with Gaussian noise in diffusion steps $[0 - T]$. At diffusion step $T$, the original signal evolves into a signal with the same dimensions as the original, but with data values following a normal distribution. In the backward process, we randomly generate signals from Gaussian noise as inputs to the diffusion model in diffusion step $T$. We then allow the diffusion model to progressively reduce noise from step $T$ to step $0$. The input for step $t$ is the output from step $t+1$. In each step $t$, we compute the KL divergence between the signal in forward step $t$ and backward step $t$, and strive to minimize the difference between them. Ultimately, signals in backward step $0$ will resemble the original signals. Once the model is trained, we can input random Gaussian noise into the diffusion model and use the backward process to generate synthetic signals. This process is considered unconditional because it imposes no constraints on how a signal should be generated from Gaussian noise. Consequently, the diffusion model learns the entire dataset distribution independently and can potentially generate any signals that reside within the feature space of the complete dataset.

\begin{figure}%[htbp]
\centering
\includegraphics[width=\columnwidth]{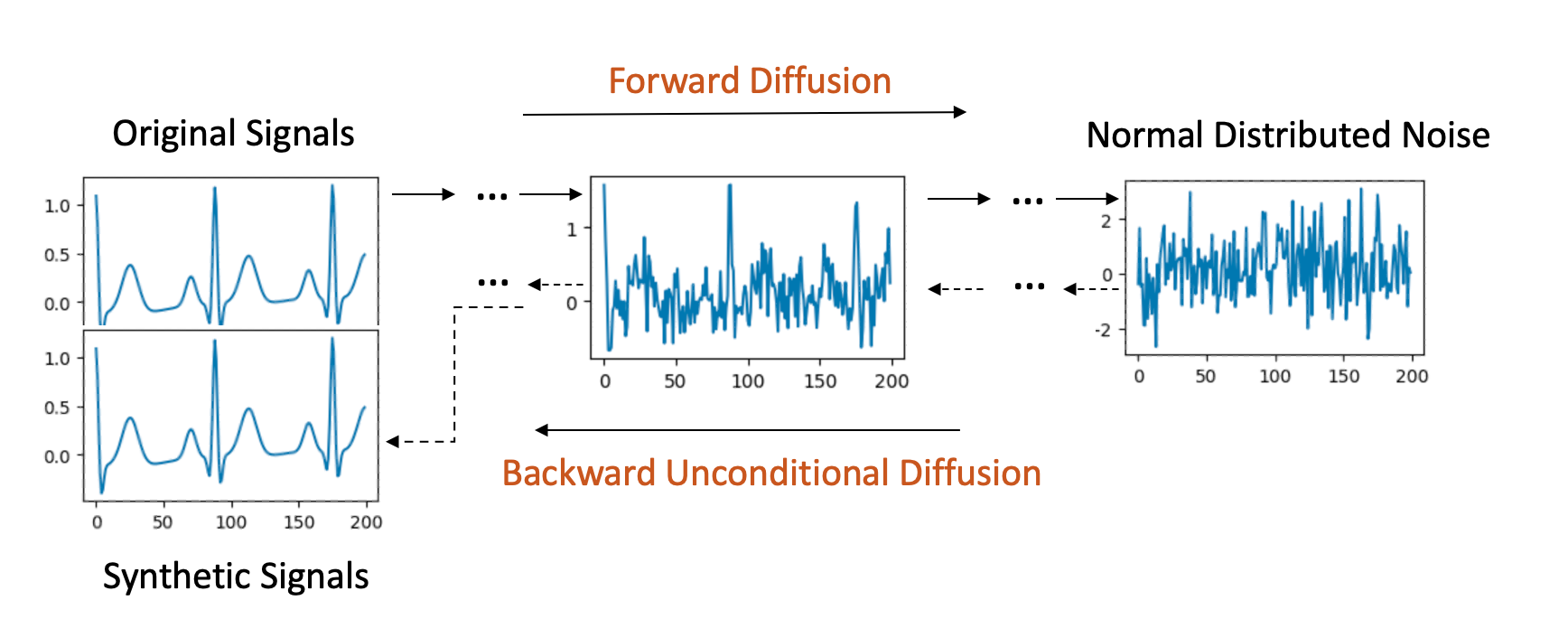}
\caption{Unconditional Diffusion process}
\label{fig:unconditional_diffusion}
\vspace{-5mm}
\end{figure}

The unconditional diffusion model employs a Markov chain-based generation process, converting data iteratively between its original form and noise. This intricate transformation is portrayed in Figure~\ref{fig:unconditional_diffusion}.

\textbf{Forward Process}: Starting with the original signal, it is incrementally perturbed with Gaussian noise over a series of diffusion steps, spanning $[0, T]$. By the end of step $T$, the resulting signal retains the dimensions of the original but its data values adopt a normal distribution.

\textbf{Backward Process}: Initiating this process, signals derived from Gaussian noise serve as inputs at diffusion step $T$. As the model retraces the steps back to $0$, it methodically diminishes the noise. Each step $t$ consumes the previous step's output ($t+1$) as its input. A crucial aspect during this phase is the evaluation of the KL divergence between signals at the corresponding steps in both the forward and backward processes. The objective is to minimize this divergence. When the backward process culminates at step $0$, the signals generated should closely mirror the original ones.

\textbf{Signal Generation}: Post-training, the model is equipped to accept random Gaussian noise. By invoking the backward process, it can craft synthetic signals. This procedure is dubbed "unconditional" due to the absence of stipulations on the signal generation from the noise. Such a design empowers the diffusion model to assimilate the dataset's entire distribution, endowing it with the capability to potentially produce any signal within the dataset's feature space.

\subsection{Label Conditional Diffusion Models}

% The label-conditional diffusion process parallels the unconditional one, with the key distinction that each input datum is coupled with a scalar label in both the forward and backward procedures. As illustrated in Fig.~\ref{fig:labelconditionalprocess}, the original signals are matched with their respective labels during the forward process. In the U-Net model (see section~\ref{sec:unet}), each residual block is injected with the scalar label and the current diffusion timestep, in the form of an embedding. The backward process involves the generation of normally distributed noise, paired with a given label, which are then fed into the diffusion model. At each diffusion step, the KL divergence between the forward and backward process signals is computed with the aim to minimize their discrepancy. Ultimately, the approach trains a diffusion model that not only comprehends the data distribution of an entire dataset but also generates synthetic signals belonging to a specific class.

Label-conditional diffusion models extend the framework of their unconditional counterparts by integrating scalar labels with each input datum. This inclusion of labels not only shapes the diffusion process but also allows for more targeted synthesis of signals, as elaborated below.

\begin{figure}%[htbp]
\centering
\includegraphics[width=\columnwidth]{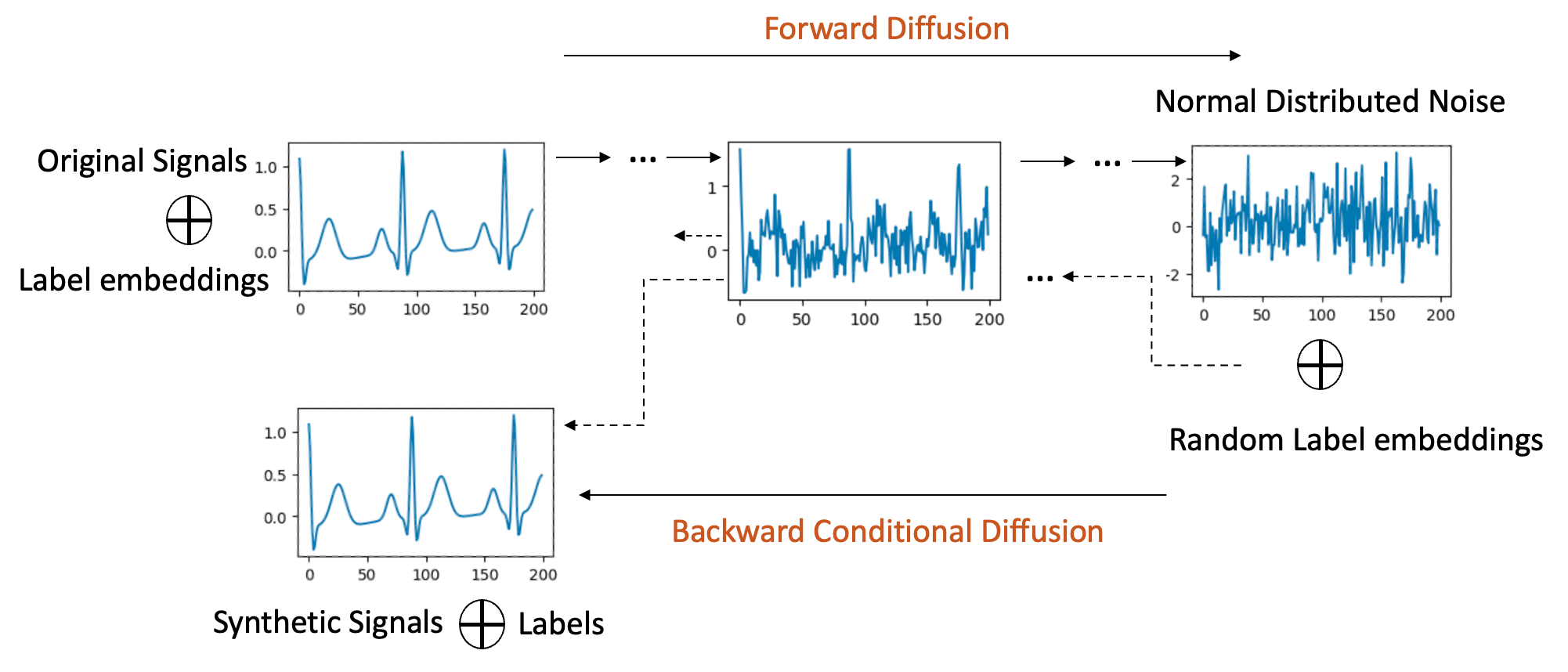}
\caption{Label Conditional Diffusion process}
\label{fig:labelconditionalprocess}
\vspace{-5mm}
\end{figure}

\textbf{Forward Process with Labels}: In this process, as depicted in Fig.~\ref{fig:labelconditionalprocess}, original signals are systematically associated with their corresponding labels. Within the U-Net architecture (detailed in section~\ref{sec:unet}), each residual block is enriched with both the scalar label and the ongoing diffusion timestep, leveraging an embedding technique.

\textbf{Backward Process with Labels}: Here, the diffusion model ingests noise, drawn from a normal distribution, in tandem with a designated label. As the model progresses through the diffusion steps, there's a persistent focus on quantifying and minimizing the KL divergence between the signals emerging from the forward and backward processes.

\textbf{Synthetic Signal Generation}: The culmination of this methodology is a trained diffusion model possessing dual capabilities. It is not only attuned to the holistic data distribution of the dataset but is also adept at crafting synthetic signals pertinent to a delineated class.

\subsection{Signal Conditional Diffusion Models}
% The signal conditional diffusion process is illustrated in Fig.~\ref{fig:sig_conditional}. This procedure contrasts with the label conditional diffusion process, in that the signal conditions are only incorporated during the backward diffusion process. In the forward diffusion process, Gaussian noise is progressively appended to the original signal from timestep $0-T$, culminating in normally distributed noise at timestep $T$. During the backward process, a signal that has been distorted serves as a conditional input. This could be an original signal that has been contaminated with random noise or artifacts, or a downsampled signal conforming to the original signal's shape. This conditional input is concatenated with normally distributed noise, which is then passed through a convolution layer to match the original signal's shape. Subsequently, the backward process methodically denoises the signal to achieve a resemblance with the original input. 

\begin{figure}%[htbp]
\centering
\includegraphics[width=\columnwidth]{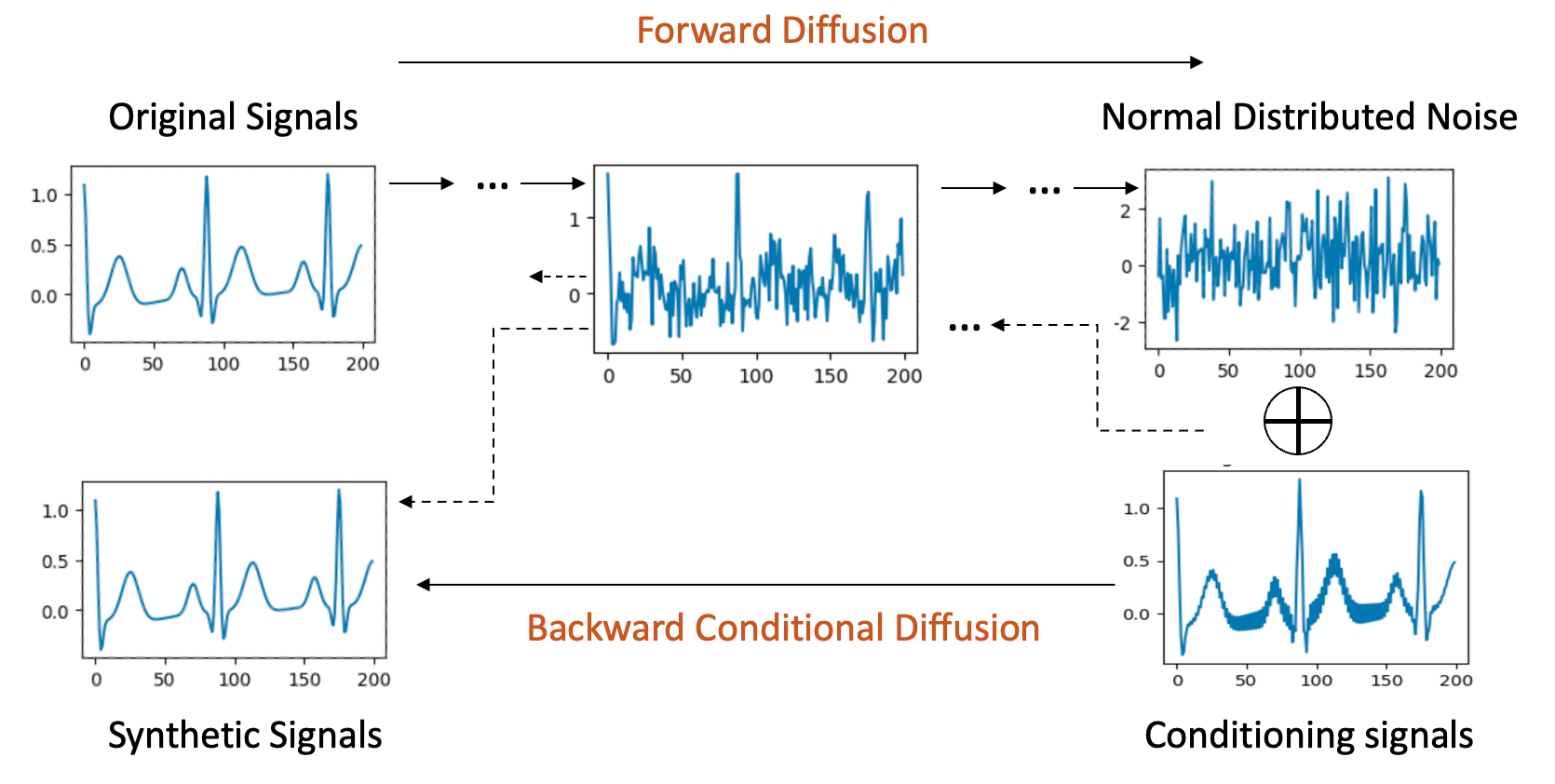}
\caption{Signal Conditional Diffusion process}
\label{fig:sig_conditional}
\vspace{-3mm}
\end{figure}

Signal-conditional diffusion models, visualized in Fig.~\ref{fig:sig_conditional}, introduce a nuanced methodology where signal conditions play a pivotal role exclusively during the backward diffusion phase, differentiating it from label-conditional approaches.

\textbf{Forward Process}: The forward diffusion process in the case of signal conditioning is the same as the original, unconditional diffusion.

\textbf{Backward Diffusion with Signal Conditioning}: For the backward phase, a perturbed signal forms the conditional input, which could stem from an original signal sample tainted by noise, artifacts, or even be a downsampled version mirroring the original signal's dimensions. This conditional signal is amalgamated with noise drawn from a normal distribution. Post this combination, a convolutional layer refines it to align with the original signal's structure. The remainder of the backward process strives to cleanse the noise and produce a clean signal resembling the original signal it was seeded with.

\subsection{U-Net Architecture}\label{sec:unet}

\begin{figure*}
\centering
\includegraphics[width=.8\textwidth]{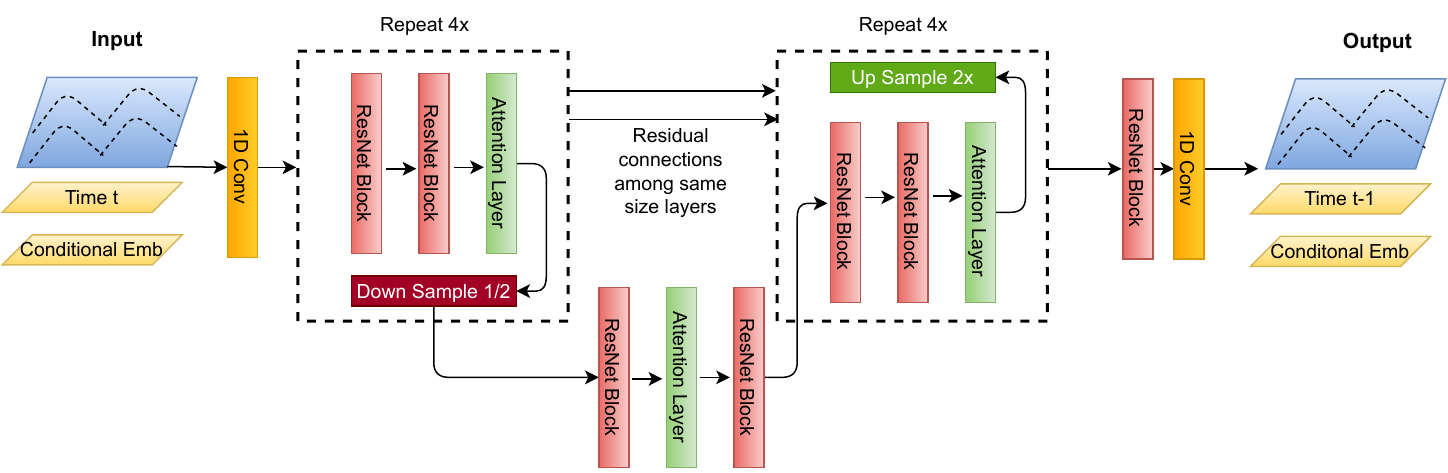}
\caption{Description of the U-Net architecture for signals with skip connections}
\label{fig:unet}
\end{figure*}

The U-Net architecture (see Figure~\ref{fig:unet}) is used in each diffusion process pipeline. We modify the model depicted in the work~\cite{saharia2022image} to fit for time-series signals instead of $NxN$ images. The signals in time step $\mathbf{x_{t}}$ are concatenated with the time step embeddings and other conditional embeddings (low-quality signals or labels) input to the U-Net model, and the U-Net model generates signals in time $t-1$. We repeat this process from $t = T$ till $t= 0$ to train the U-Net model. 

\section{Experimental Results}
\label{sec:experiments}
This section delves into the various methodologies employed by diffusion models in the synthesis of biomedical signals. We've partitioned our approach into three categories: unconditional, label-conditional, and signal-conditional diffusion processes. Our qualitative and quantitative evaluations underscore the efficacy of the generated synthetic data. We also benchmark our findings against state-of-the-art methods, underscoring the advantages of our model and pinpointing areas ripe for further refinement. We aim to demonstrate that diffusion models are promising candidates for crafting high-caliber biomedical signals, potentially transforming myriad biomedical arenas.

\subsection{Datasets}
Our model's performance is gauged across three datasets: a simulated one, the UniMiB human activity recognition (HAR) dataset~\cite{app7101101}, and the MIT-BIH Arrhythmia Database~\cite{moody2001impact, goldberger2000physiobank}.

\begin{itemize}
    \item \textbf{Simulated Dataset}: Comprising synthetic signals with patterns like bell, funnel, and cylinder shapes, this dataset contains five classes each defined by attributes such as amplitude and pattern variances. Each signal has 512 timesteps and is unidimensional. Its primary purpose is to assess the diffusion model's adaptability on simplified data.

    \item \textbf{UniMiB Dataset}~\cite{app7101101}: Derived from smartwatches, this dataset has nine human activity classes with each signal capturing 151 timesteps across three acceleration dimensions. Adapted to our U-Net architecture, signals have been resized to 128 timesteps. The training set contains \textbf{6055} samples, with class distributions that peak at 1572 and trough at 119 samples per class. The test set has \textbf{1524} samples, ranging from 32 to 413 samples per class, highlighting the dataset's imbalance.

    \item \textbf{MIT-BIH Arrhythmia Dataset}: Featuring 48 snippets of ambulatory ECG recordings spanning half an hour each from 47 subjects across five heart conditions~\cite{moody2001impact, goldberger2000physiobank}. The samples, originally recorded at 125Hz, have been adjusted to 144 in length for U-Net compatibility. The training set has \textbf{87554} samples, with the majority class having 72471 samples and the smallest class having 641. The test set includes \textbf{21892} samples, ranging from 162 to 18118 samples per class, again underlining the dataset's imbalance.
\end{itemize}

For an in-depth exploration of the datasets, kindly refer to the Appendix.

\begin{figure*}%[htbp]
  \centering
  \begin{subfigure}[b]{\columnwidth}
    \includegraphics[width=\textwidth]{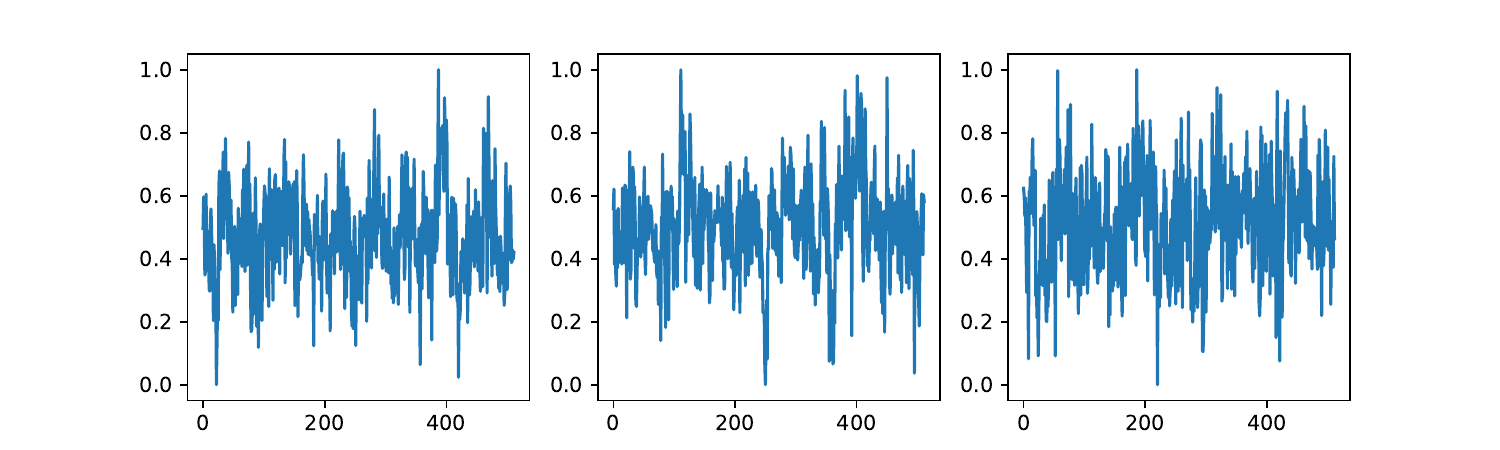}
    \caption{Simulated real class-0}
    \label{fig:subfig1}
  \end{subfigure}
  ~
  \begin{subfigure}[b]{\columnwidth}
    \includegraphics[width=\textwidth]{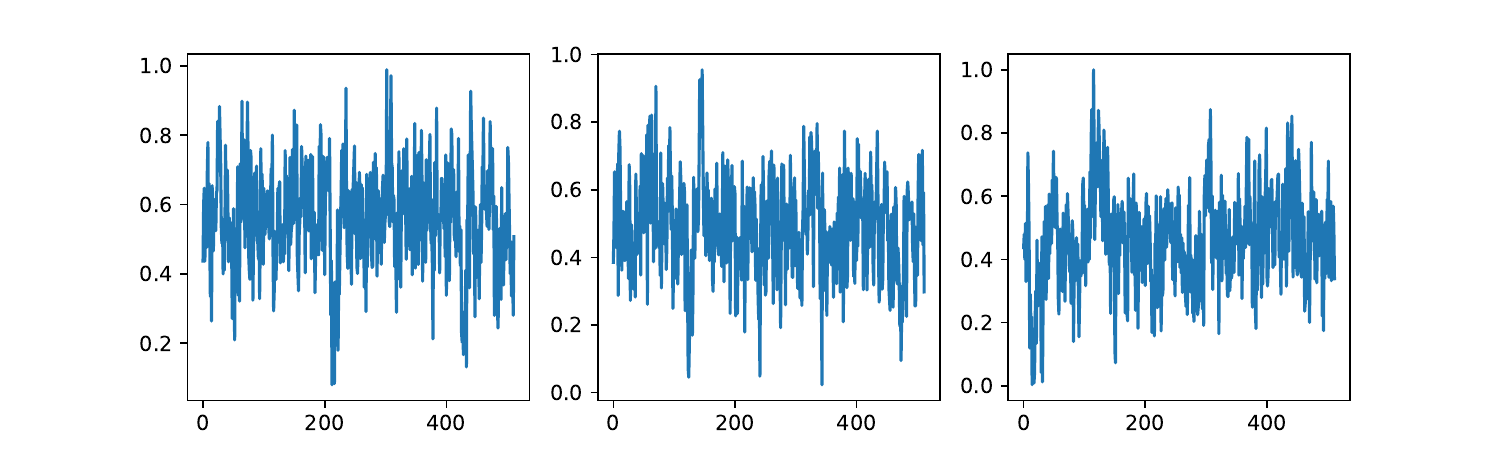}
    \caption{Simulated synthetic class-0}
    \label{fig:subfig2}
  \end{subfigure}

  \medskip

  \begin{subfigure}[b]{\columnwidth}
    \includegraphics[width=\textwidth]{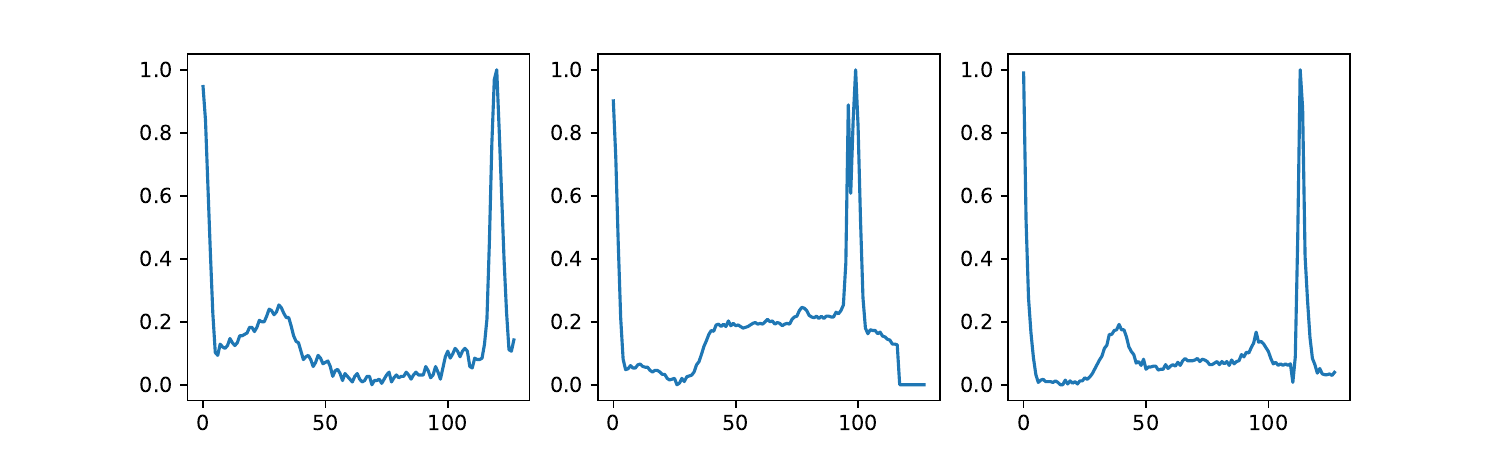}
    \caption{MITBIH real class-0}
    \label{fig:subfig3}
  \end{subfigure}
  ~
  \begin{subfigure}[b]{\columnwidth}
    \includegraphics[width=\textwidth]{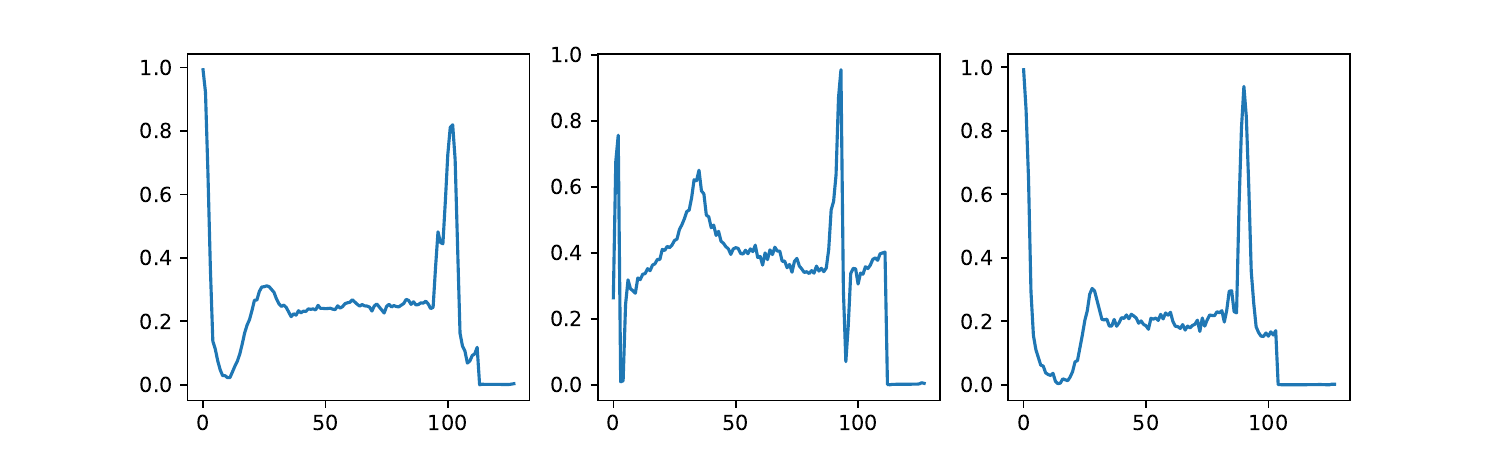}
    \caption{MITBIH synthetic class-0}
    \label{fig:subfig4}
  \end{subfigure}

  \medskip

  \begin{subfigure}[b]{\columnwidth}
    \includegraphics[width=\textwidth]{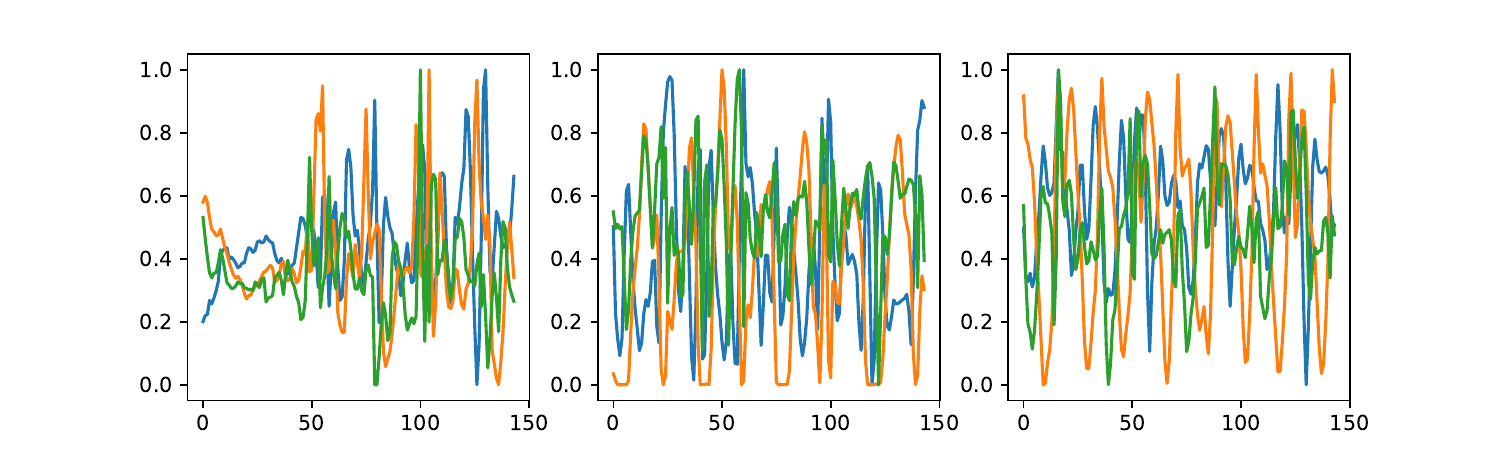}
    \caption{UNIMIB real GoingDownS}
    \label{fig:subfig5}
  \end{subfigure}
  ~
  \begin{subfigure}[b]{\columnwidth}
    \includegraphics[width=\textwidth]{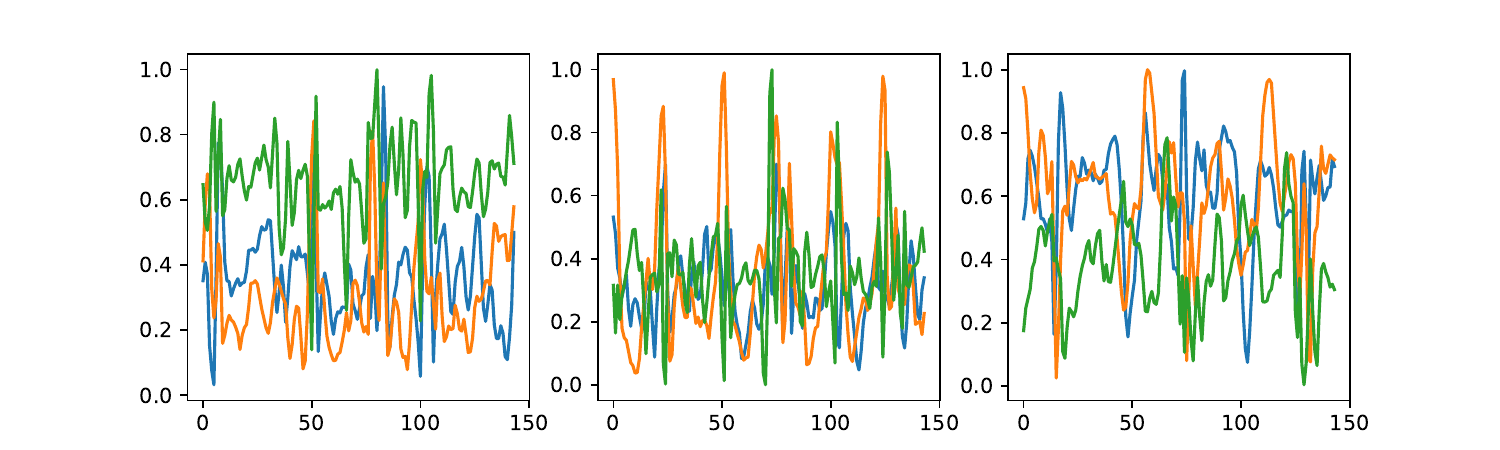}
    \caption{UNIMIB synthetic GoingDownS}
    \label{fig:subfig6}
  \end{subfigure}

  \caption{Raw signals comparison. The left column shows real raw signals. The right column shows synthetic raw signals.}
  \label{fig:rawsignals}
\end{figure*}

\begin{figure*}%[htbp]
    \centering
    \begin{subfigure}[b]{0.32\textwidth}
        \includegraphics[width=\textwidth]{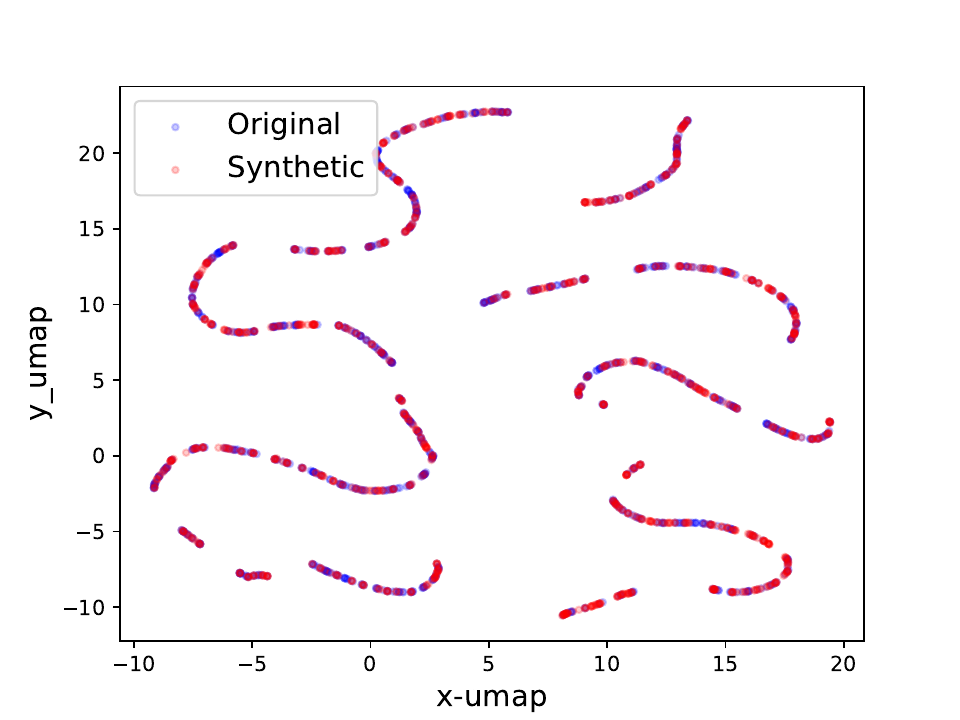}
        \caption{Simulated dataset Class 4}
    \end{subfigure}
    \hfill
    \begin{subfigure}[b]{0.32\textwidth}
        \includegraphics[width=\textwidth]{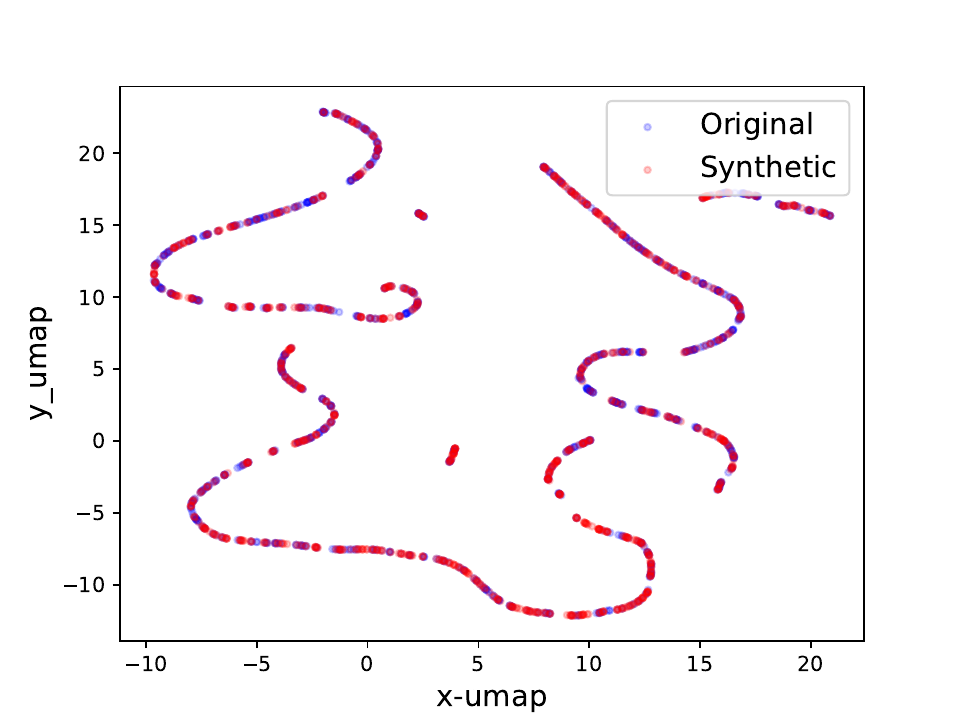}
        \caption{MITBIH dataset Class 4}
    \end{subfigure}
    \hfill
    \begin{subfigure}[b]{0.32\textwidth}
        \includegraphics[width=\textwidth]{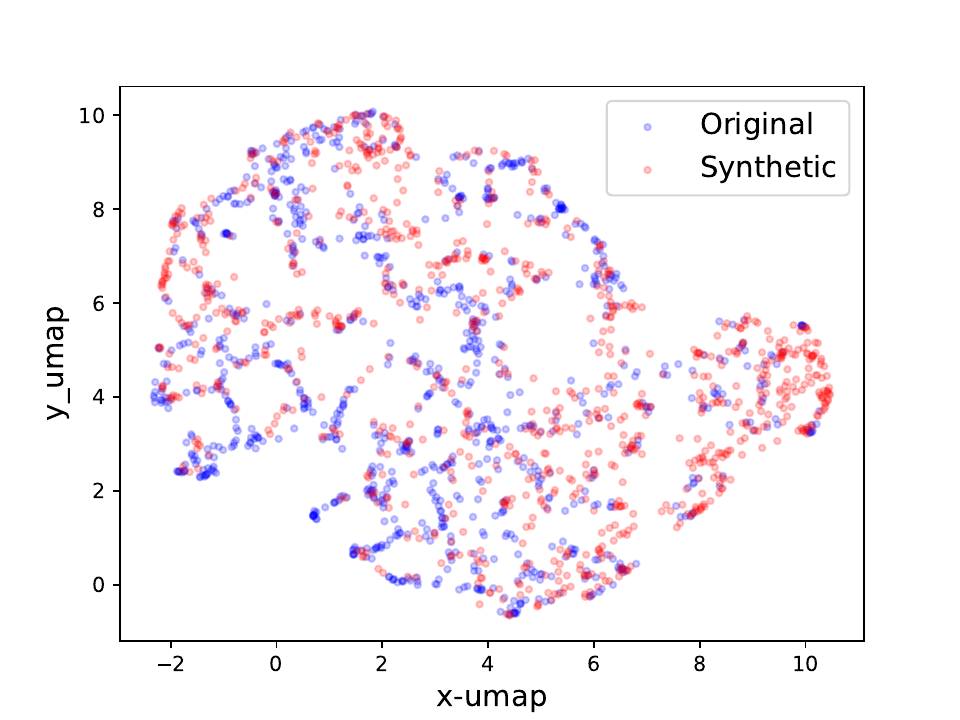}
        \caption{UNIMIB dataset Walking}
    \end{subfigure}
    \caption{The real and synthetic data UMAP projection on three classes of three datasets. Each red dot represents one original signal after the dimension reduction, whereas a blue dot represents one synthetic signal. From the graphs, we can see that a set of synthetic signals are having similar distribution as a set of real signals in the 2D UMAP projections graphs.}
    \label{fig:three_umaps}
\end{figure*}

\subsection{Visualization of Raw Signals}
To assess the fidelity of synthetic signals visually, we present a comparative plot between several real and synthetic signals. For continuity, discrete signal values at each sampling interval are interconnected. Figure~\ref{fig:rawsignals} illustrates a set of both real and synthetic signals derived from three distinct datasets. An immediate examination reveals the capability of our diffusion model in crafting synthetic signals that closely mirror the genuine signals.

\subsection{Projection Through Dimension Reduction}
For each class in every dataset, an unconditional diffusion model is trained. The UMAP projection of synthetic signals in relation to the original ones for select data classes is depicted in Figure~\ref{fig:three_umaps}. Extended visualizations are accessible in the provided source code repository. When scrutinized, it becomes evident that even for signals of considerable length (e.g., 512 timesteps), our diffusion model adeptly recognizes and replicates the intricate signal patterns. Moreover, the synthetic signals span the entire feature spectrum inhabited by the genuine signals.

\subsection{Similarity Metrics and Results}
The label-conditional diffusion model not only emulates the synthetic signal generation prowess of the unconditional model but also provides a guided synthesis tailored for specific classes. While the raw signals and UMAP projections closely resemble the ones in Figures~\ref{fig:rawsignals} and~\ref{fig:three_umaps}, the main advantage lies in the training efficiency. A singular label-conditional diffusion model suffices for a multi-class dataset, in contrast to the multiple models required by the unconditional counterpart for each class. Intriguingly, when it comes to sparsely represented data classes, the label-conditional model potentially outperforms the unconditional one. This edge is attributed to its capacity to generalize patterns across the dataset and utilize this knowledge for class-specific synthesis.

To underscore the fidelity of signals generated by our diffusion models, we calculate similarity scores across diverse signal classes. The results are cataloged in Table~\ref{tbl:similarity_scores}. Our BioDiffusion model's outputs closely align with real signals, surpassing the fidelity of other cutting-edge techniques.

\noindent
\textbf{Quantitative Metrics}:
\begin{itemize}
    \item \textbf{Wavelet Coherence}: A measure to assess common oscillations between two time-series across specific frequencies over time. Values oscillate between 0 and 1, with 1 denoting impeccable coherence. Due to its proficiency with non-stationary signals, it's a vital tool for evaluating evolving spectral content~\cite{li2022tts}.
    \item \textbf{Discriminative Score}: Conceived as a quantitative metric to juxtapose sequences from real and generated datasets. It employs a 2-layer LSTM classifier trained to segregate the two datasets. The classification error on a separate test set offers an objective similarity evaluation~\cite{yoon2019time}.
\end{itemize}

\noindent
\textbf{Baseline Techniques}: 
\begin{itemize}
    \item \textbf{C-RNN-GAN}: A pioneering GAN-based solution for sequential data synthesis using two-layer LSTM for both generator and discriminator~\cite{mogren2016c}.
    \item \textbf{RCWGAN}: An enhanced version of C-RNN-GAN with conditional data input for controlled generation~\cite{esteban2017real}.
    \item \textbf{TimeGAN}: A groundbreaking GAN framework that harnesses a latent space for time-series synthesis, augmented with both supervised and unsupervised losses~\cite{yoon2019time}.
    \item \textbf{SigCWGAN}: Enhances the GAN process with conditional data and the Wasserstein loss for stable training~\cite{ni2020conditional}.
    \item \textbf{TTS-GAN}: A novel transformer-centric GAN model focusing on high-fidelity single-class time-series generation~\cite{li2022tts}.
    \item \textbf{TTS-CGAN}: An iterative version of TTS-GAN introducing a label conditional transformer GAN, facilitating multi-class synthesis through a singular model~\cite{li2022ttscgan}.
\end{itemize}

\begin{table}
\caption{Comparison scores of real and synthetic data generated by different state-of-the-art time-series generation models.}
\label{tbl:similarity_scores}
\resizebox{\columnwidth}{!}{
\begin{tabular}{ccccc}
\hline
\multicolumn{5}{c}{Wavelet Coherence score (the higher the better)}                                                                                                  \\ \hline
Models & SittingDown & Jumping & Non-Ectopic & FusionBeats \\ 
\hline
C-RNN-GAN & 41.10 & 40.29 & 30.44 & 25.51 \\
RCWGAN & 39.90 & 38.85 & 29.72 & 22.97 \\
TimeGAN & 40.45 & 39.42 & 31.55 & 21.98 \\
SigCWGAN & 41.60 & 41.02 & 31.36 & 22.87 \\
TTS-GAN & 43.92 & 47.64 & 45.30 & 55.64 \\
TTS-CGAN & 45.07 & 47.64 & 47.79 & 58.34 \\
BioDiffusion & \textbf{78.17} & \textbf{90.30} & \textbf{89.30} & \textbf{91.81} \\
\hline
\multicolumn{5}{c}{Discriminative score (the lower the better)} \\ 
\hline
Models & SittingDown & Jumping & Non-Ectopic & FusionBeats \\
C-RNN-GAN & 0.308 & 0.304 & 0.189 & 0.493 \\
RCWGAN & 0.294 & 0.311 & 0.483 & 0.499 \\
TimeGAN & 0.261 & 0.217 & 0.464 & 0.312 \\
SigCWGAN & 0.310 & 0.308 & 0.413 & 0.491 \\
TTS-GAN & 0.294 & 0.167 & 0.107 & 0.380 \\
TTS-CGAN & 0.191 & \textbf{0.057} & 0.162 & 0.261 \\
BioDiffusion & \textbf{0.126} & 0.121 & \textbf{0.159} & \textbf{0.231} \\
\hline
\end{tabular}
}
\end{table}

\subsection{Utility of Synthetic Signals in Addressing Class Imbalance}
To explore the potential of synthetic signals in rectifying class imbalance issues, we constructed a classification experiment centered around the MIT-BIH dataset. This dataset, while demonstrating commendable overall accuracy, manifests stark class imbalances, often disadvantaging minority classes in terms of precision and recall.

\textbf{Experimental Setup:}
Initially, a 1D-CNN classification model was trained on the MIT-BIH training set. Given the inherent class disparities, minority classes recorded suboptimal precision and recall metrics, adversely influenced by dominant classes. Subsequently, we employed our label diffusion model to generate synthetic signals for each class. These synthetic signals were incorporated into the training set, striving to alleviate the dataset's imbalance. The identical classification model was retrained and then assessed using the same genuine test set. For a comprehensive evaluation, our method's performance was juxtaposed against traditional resampling techniques and other generative models employed for signal synthesis.

\textbf{Results and Analysis:}
As presented in Table~\ref{tbl:mitbih_classify}, synthetic signals crafted using our BioDiffusion model not only enhanced the training set but also significantly bolstered the F1-score for the detection of minority classes. In contrast, signals synthesized by models like RCWGAN and C-RNN-GAN led the downstream classifier to a biased classification—predominantly towards the majority class (Non-Ectopic Beats), effectively nullifying the F1-score for other classes. It is pivotal to note that during these evaluations, the real test set remained untouched and unseen throughout all generative model training phases.

%introduction of similarity scores
% wavelet coherence score
% discriminative score
% predictive score

%introduction of baseline models
%TTS-GAN
%...

\begin{table}
\caption{Per-class F1-scores for MIT-BIH classification, using synthetic data to mitigate class imbalance. Abbreviations: N-Ect = Non-Ectopic Beats, S-Ect = Superventricular Ectopic Beats, Vent = Ventricular Beats, Unk = Unknown Beats, Fus = Fusion Beats.}
\label{tbl:mitbih_classify}
\resizebox{\columnwidth}{!}{
\begin{tabular}{l|lllll|l}
\hline
             & N-Ect & S-Ect         & Vent          & Unk           & Fus           & Average        \\ \hline
Imbalanced   & \textbf{0.97}  & 0.25          & 0.75          & 0.38          & 0.89          & 0.648          \\
Re-sampling  & 0.50  & 0.65          & 0.64          & 0.81          & 0.85          & 0.69           \\
% C-RNN-GAN    & 0.33  & 0             & 0             & 0             & 0             & 0.066          \\
% RCWGAN       & 0.33  & 0             & 0             & 0             & 0             & 0.066          \\
TimeGAN      & 0.60  & 0.48          & 0.75          & 0.48          & 0.93          & 0.648          \\
SigCWGAN     & 0.59  & 0.60          & 0.80          & 0.58          & 0.93          & 0.7            \\
TTS-GAN      & 0.60  & 0.77          & 0.75          & 0.60          & 0.91          & 0.726          \\
TTS-CGAN     & 0.66  & 0.78          & 0.77          & 0.85          & 0.93          & 0.798          \\
BioDiffusion & 0.73  & \textbf{0.79} & \textbf{0.86} & \textbf{0.84} & \textbf{0.95} & \textbf{0.834} \\
\hline
\end{tabular}
}
\end{table}

\begin{figure}%[htbp]
    \centering
    \includegraphics[trim=2in 0in 2in 0in, clip, width=\columnwidth]{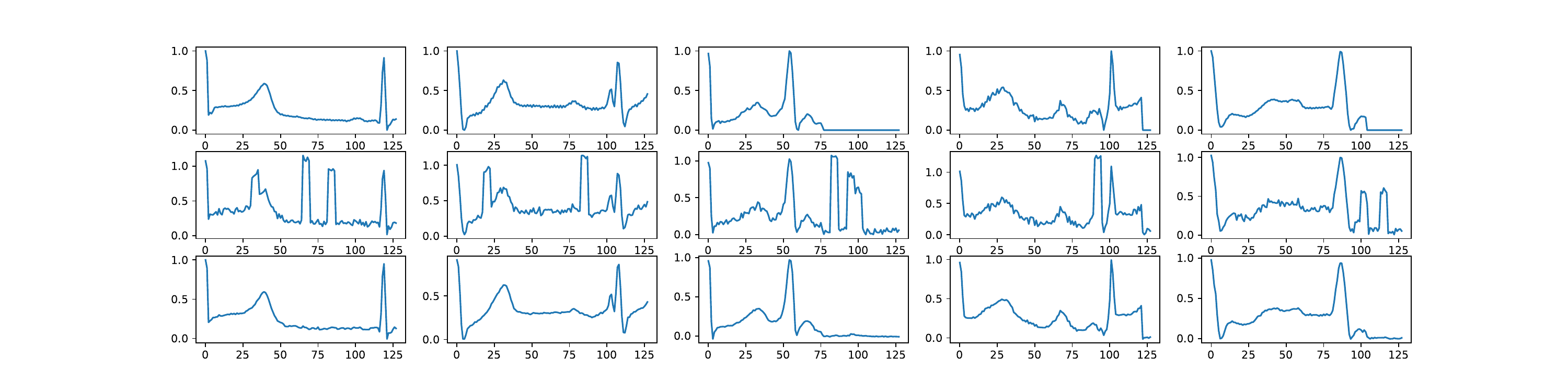}
    \caption{Example signal denoising results. First row: real signals. Second row: signals with noise. Third row: denoised signals using BioDiffusion.}
    \label{fig:signal_denoising}
    \vspace{-4mm}
\end{figure}

% \subsection{Signal Imputation}
% Signal imputation is another task that BioDiffusion can handle. Oftentimes, the collected signals may contain some missing values. We can use BioDiffusion to fill in those blanks. Fig.~\ref{fig:signal_imputation} shows a few examples of signal imputation. The first row shows the original signals. The second row shows the same signals with some randomly missing values (values set to zero). We use them as signal conditions input to the diffusion model. The third row shows the reconstructed signals. We can see that the synthetic signals fill in the blanks and are very similar to the original signals. 

\begin{figure}%[htbp]
    \vspace{-3mm}
    \centering
    \includegraphics[trim=2in 0in 2in 0in, clip,width=\columnwidth]{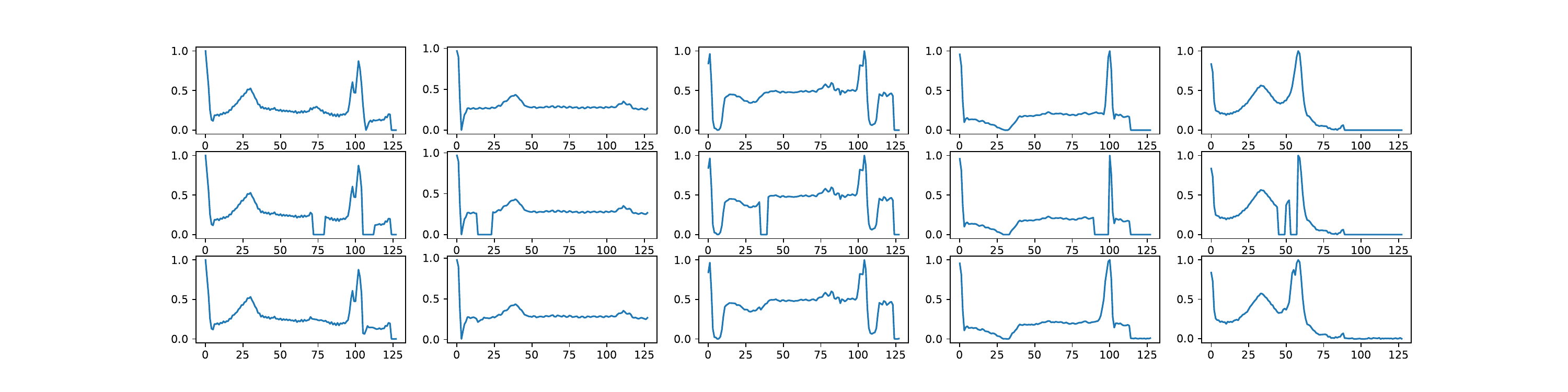}
    \caption{Example signal imputation results. First row: real signals. Second row: signals with random blanks. Third row: imputed signals by BioDiffusion model.}
    \label{fig:signal_imputation}
\end{figure}

\subsection{BioDiffusion for in Biomedical Signal Denoising, Imputation, and Upsampling}

In time-series signal collections, three predominant noise types are frequently encountered: thermal noise, electrode contact noise, and motion artifacts. Thermal noise arises from the thermal agitation of electrons causing voltage or current fluctuations. Electrode contact noise stems from the changing electrical characteristics between electrodes and surfaces leading to signal baseline fluctuations. Motion artifacts, on the other hand, are sudden spikes in signals due to physical disturbances like movement, unrelated to the actual biological activity being measured. These artifacts and noise types 
%(illustrated in the appendix as Fig.\ref{fig:signal_noise}), 
challenge the robustness of signal processing techniques. Leveraging BioDiffusion, we successfully denoised signals by taking the MIT-BIH dataset, adding artificial noise, and using it as an input for the diffusion model. 
% The results are showcased in Fig.~\ref{fig:signal_denoising}, where the synthetic signals closely resemble the original signals from the test set.

Furthermore, BioDiffusion efficiently handles signal imputation tasks. Missing values in collected signals can be interpolated using our model, resulting in reconstructed signals that are impressively close to the original signals, as displayed in Fig.~\ref{fig:signal_imputation}.

Differing sampling rates across biomedical signals, due to equipment variations, necessitate resampling techniques. Traditional upsampling methods, while functional, often fail to capture intricate relationships among signal features. This problem is addressed with our signal conditional diffusion model designed for signal upsampling, resulting in high-resolution signals that are almost indistinguishable from the originals. 
% Fig~\ref{fig:signal_sr_lr} in the appendix gives a visual comparison of the efficacy of our model in this domain.

A notable application of BioDiffusion lies in the generation of individualized signals. Scarcity of data samples from individual subjects can be a bottleneck for certain machine learning applications. However, our approach allows a diffusion model to be trained on diverse signals, which is then fine-tuned using signals from an individual subject. This method generates synthetic signals that retain the unique patterns of the subject, enabling the expansion of subject-specific datasets. 
% Such synthetic signal generation can be visualized in Fig.~\ref{fig:single_generation}.

For more visual examples of the output of BioDiffusion in upsampling and personalized signal generation, please see the Appendix.

\section{Conclusion}
\label{sec:conclusion}
In conclusion, the proposed BioDiffusion model is a novel and versatile probabilistic model specifically designed for generating synthetic biomedical signals. Our model offers a comprehensive solution for various generation tasks, including unconditional, label-conditional, and signal-conditional generation, which makes it a valuable tool for biomedical signal synthesis. We evaluated the quality of the generated signals using qualitative and quantitative assessments and demonstrated the effectiveness and accuracy of the BioDiffusion model in producing high-quality synthetic biomedical signals. Compared to state-of-the-art time-series synthesis models, our BioDiffusion model consistently outperforms its counterparts, showcasing its superiority and robustness in biomedical signal generation. The model's versatility and adaptability have the potential to significantly contribute to the advancement of biomedical signal processing techniques, opening up new possibilities for improved research outcomes and clinical applications.

\clearpage

\bibliographystyle{plain}
\bibliography{paperdraft}

\newpage
\onecolumn
\appendix

\section{Dataset details}
More of the dataset details is in Table~\ref{tb:datasets}

\begin{table}[ht]
\caption{Dataset details}
\label{tb:datasets}
\centering
\begin{tabular}{|l|l|l|l|l|l|l|l|}
\hline
Dataset                       & Signal type                                                                       & N Channels & Type  & Total samples & Classes & Class ratio                                                                       & Sample length \\ \hline
\multirow{2}{*}{Simulated} & \multirow{2}{*}{\begin{tabular}[c]{@{}l@{}}Simulated \\ signals\end{tabular}}     & 1          & Train & 20000         & 5                & 1:1:1:1:1                                                                            & 512           \\ \cline{3-8} 
                                   &                                                                                   & 1          & Test  & 2000          & 5                & 1:1:1:1:1                                                                            & 512           \\ \hline
\multirow{2}{*}{UNIMIB}            & \multirow{2}{*}{\begin{tabular}[c]{@{}l@{}}Accelerometer \\ signals\end{tabular}} & 3          & Train & 6055          & 9                & \begin{tabular}[c]{@{}l@{}}119:169:1394:\\ 1572:737:600:\\ 1068:228:168\end{tabular} & 128           \\ \cline{3-8} 
                                   &                                                                                   & 3          & Test  & 1524          & 9                & \begin{tabular}[c]{@{}l@{}}34:47:344:\\ 413:184:146:\\ 256:68:32\end{tabular}        & 128           \\ \hline
\multirow{2}{*}{MITBIH}            & \multirow{2}{*}{\begin{tabular}[c]{@{}l@{}}ECG \\ signals\end{tabular}}           & 1          & Train & 87554         & 5                & \begin{tabular}[c]{@{}l@{}}72471:2223:\\ 5788:641:6431\end{tabular}                  & 144           \\ \cline{3-8} 
                                   &                                                                                   & 1          & Test  & 21892         & 5                & \begin{tabular}[c]{@{}l@{}}18118:556:\\ 1448:162:1608\end{tabular}                   & 144           \\ \hline
\end{tabular}
\end{table}
% \begin{figure}
% \centering
% \includegraphics[width=0.8\textwidth]{images/Real UniMiB UMAP Projection.pdf}
% \caption{Real UniMiB (9 classes) UMAP Projection}
% \label{fig:Real_UniMiB UMAP}
% \end{figure}

\section{Training details}
We train an unconditional diffusion model per class per dataset. The training details are as follows. 
\begin{multicols}{3}
    \noindent\textbf{Architecture}
    \begin{itemize}
        \item Base channels: 64
        \item Channel multipliers: 1, 2, 4, 8, 8 (Simulated Dataset)
        \item Channel multipliers: 1, 2, 4, 8 (UNIMIB and MITBIH)
        \item Residual blocks groups: 8
        \item Attention heads: 4
    \end{itemize}
    \columnbreak
    
    \noindent\textbf{Training}
    \begin{itemize}
        \item Optimizer: Adam
        \item Batch size: 32
        \item Learning rate: 3e-4
        \item Epochs: 100
        \item Hardware: NVIDIA RTX A5000
    \end{itemize}
    \columnbreak
    
    \noindent\textbf{Diffusion}
    \begin{itemize}
        \item Timesteps: 1000
        \item Noise schedule: cosine
        \item Loss: l1
    \end{itemize}
\end{multicols}

We train a label condition diffusion model per dataset. Each signal sample is paired with a scalar label. The training details are as follows. 
\begin{multicols}{3}
    \noindent\textbf{Architecture}
    \begin{itemize}
        \item Base dimensions: 64
        \item Channel multipliers: 1, 2, 4, 8, 8 (Simulated Dataset)
        \item Channel multipliers: 1, 2, 4, 8 (UNIMIB and MITBIH)
        \item Number classes: 5 (Simulated and MITBIH dataset) 
        \item Number classes: 9 (UNIMIB dataset)
        \item Residual blocks groups: 8
        \item Attention heads: 4
        \item Conditional drop prob: 0.5
    \end{itemize}
    \columnbreak
    
    \noindent\textbf{Training}
    \begin{itemize}
        \item Optimizer: Adam
        \item Batch size: 32
        \item Learning rate: 3e-4
        \item Epochs: 100
        \item Hardware: NVIDIA GeForce 1080
    \end{itemize}
    \columnbreak
    
    \noindent\textbf{Diffusion}
    \begin{itemize}
        \item Diffusion timesteps: 1000
        \item Noise schedule: cosine
        \item Loss: l1
    \end{itemize}
\end{multicols}

A signal conditional model is trained on a specific class of data, and synthetic signals are generated by using distorted signals as conditional inputs. The distorted signals provided to the model were not present in the training set, with the aim of assessing the model's ability to accurately restore them to their original form. Here we present several possible implementations of the model, it should be noted that these examples are not exhaustive, and the model is capable of other implementations as well.
\begin{multicols}{3}
    \noindent\textbf{Architecture}
    \begin{itemize}
        \item Base channels: 64
        \item Channel multipliers: 1, 2, 4, 8, 8 
        \item Residual blocks groups: 2
        \item Attention heads: 4
    \end{itemize}
    \columnbreak
    
    \noindent\textbf{Training}
    \begin{itemize}
        \item Optimizer: Adam
        \item Batch size: 32
        \item Learning rate: 1e-4
        \item iterations: 1000000
        \item Hardware: NVIDIA GeForce 1080
    \end{itemize}
    \columnbreak
    
    \noindent\textbf{Diffusion}
    \begin{itemize}
        \item Timesteps: 2000
        \item Noise schedule: linear
        \item Loss: l1
    \end{itemize}
\end{multicols}

\section{More visualizations about label conditional generation}

\begin{figure}[htbp]
    \centering
    \begin{subfigure}[b]{0.22\textwidth}
        \includegraphics[width=\textwidth]{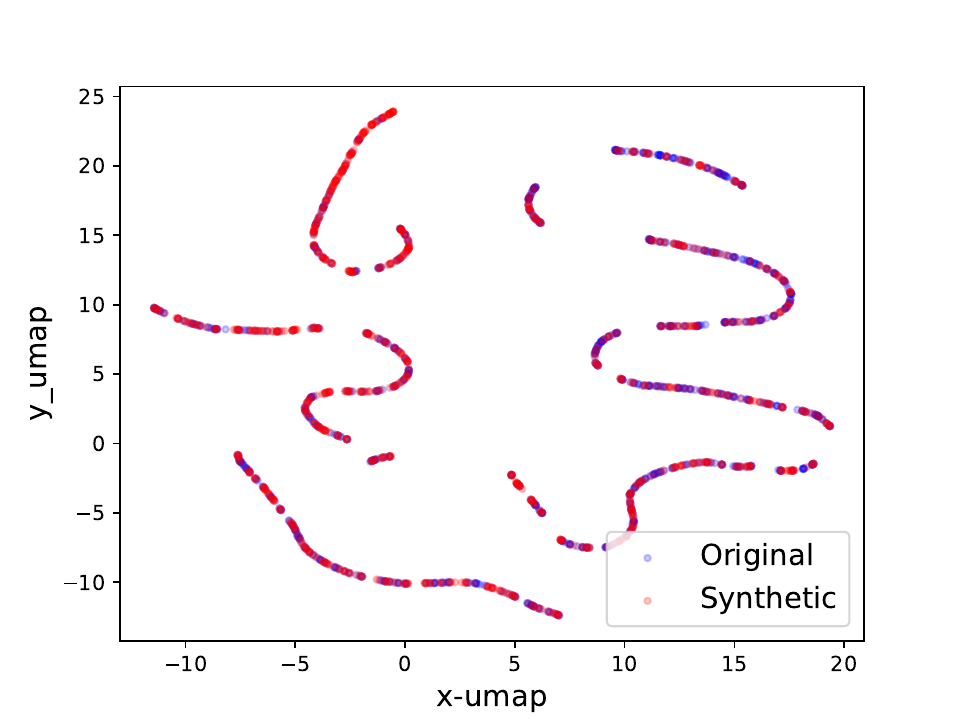}
        \caption{Class 0}
    \end{subfigure}
    \hfill
    \begin{subfigure}[b]{0.22\textwidth}
        \includegraphics[width=\textwidth]{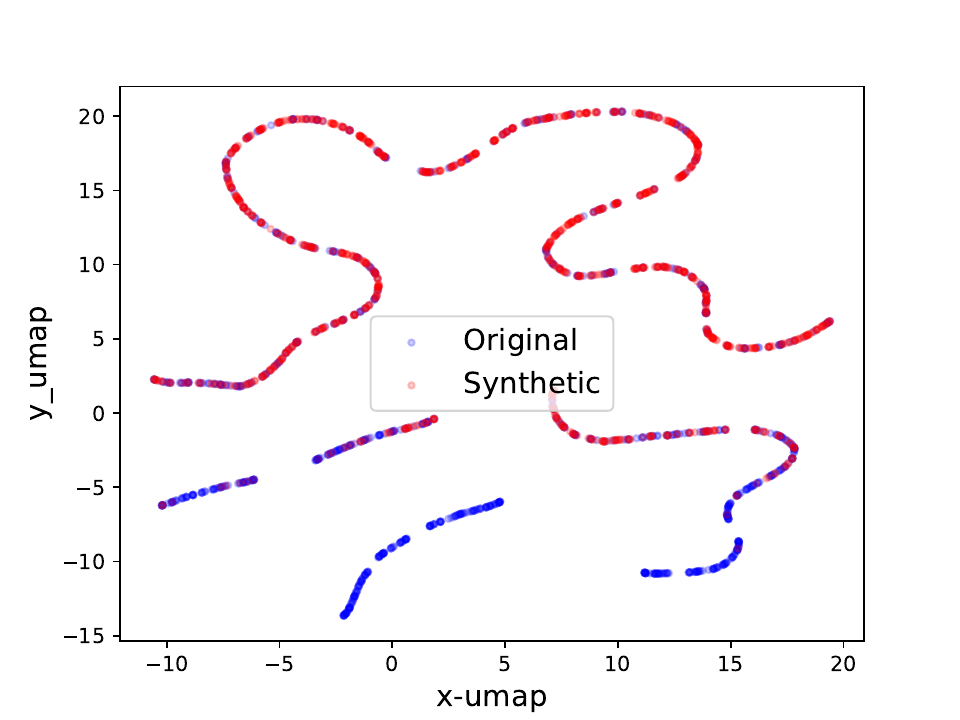}
        \caption{Class 1}
    \end{subfigure}
    \hfill
    \begin{subfigure}[b]{0.22\textwidth}
        \includegraphics[width=\textwidth]{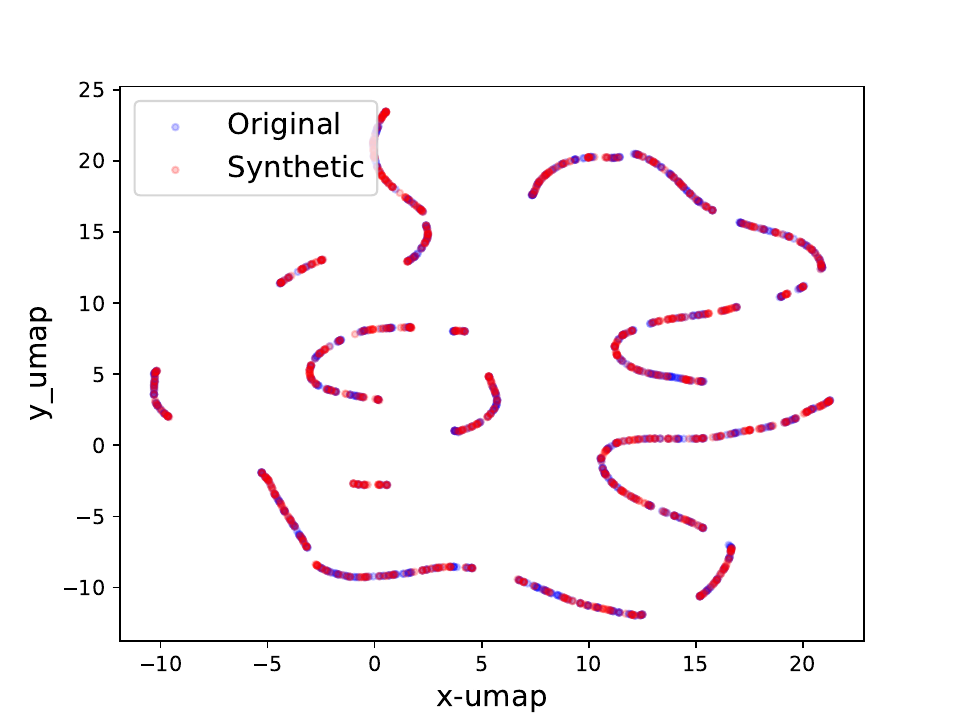}
        \caption{Class 2}
    \end{subfigure}
    \hfill
    \begin{subfigure}[b]{0.22\textwidth}
        \includegraphics[width=\textwidth]{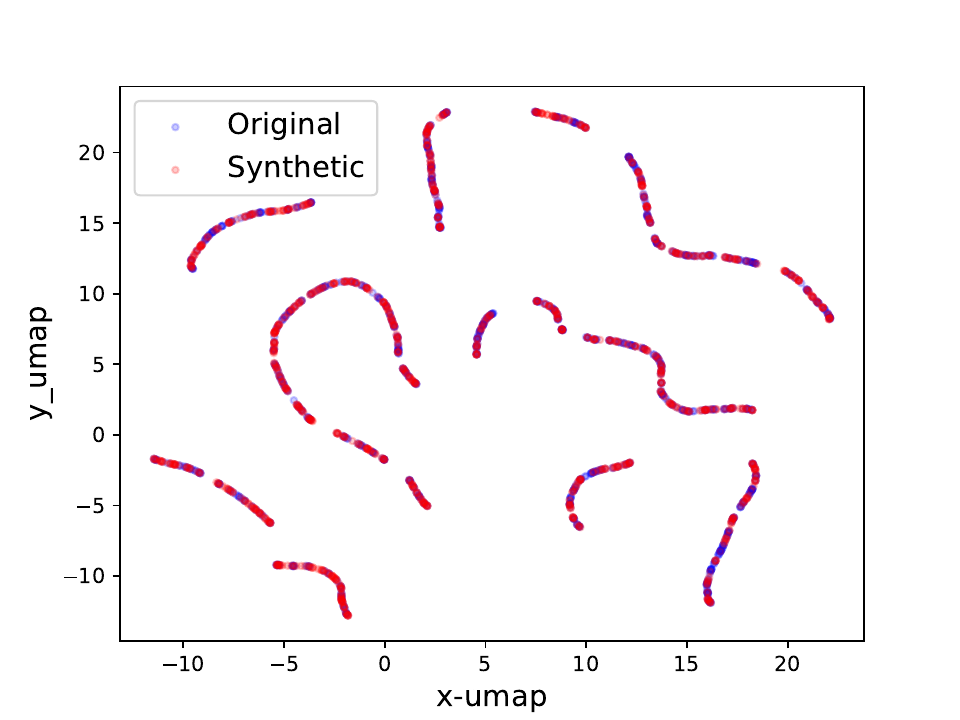}
        \caption{Class 3}
    \end{subfigure}
    \caption{Real and synthetic signal UMAP projection on selected classes of simulated dataset}
    \label{fig:simulated_umaps}
\end{figure}

\begin{figure}[htbp]
    \centering
    \begin{subfigure}[b]{0.22\textwidth}
        \includegraphics[width=\textwidth]{images/umaps/unimib_Walking_umap.pdf}
        \caption{Walking}
    \end{subfigure}
    \hfill
    \begin{subfigure}[b]{0.22\textwidth}
        \includegraphics[width=\textwidth]{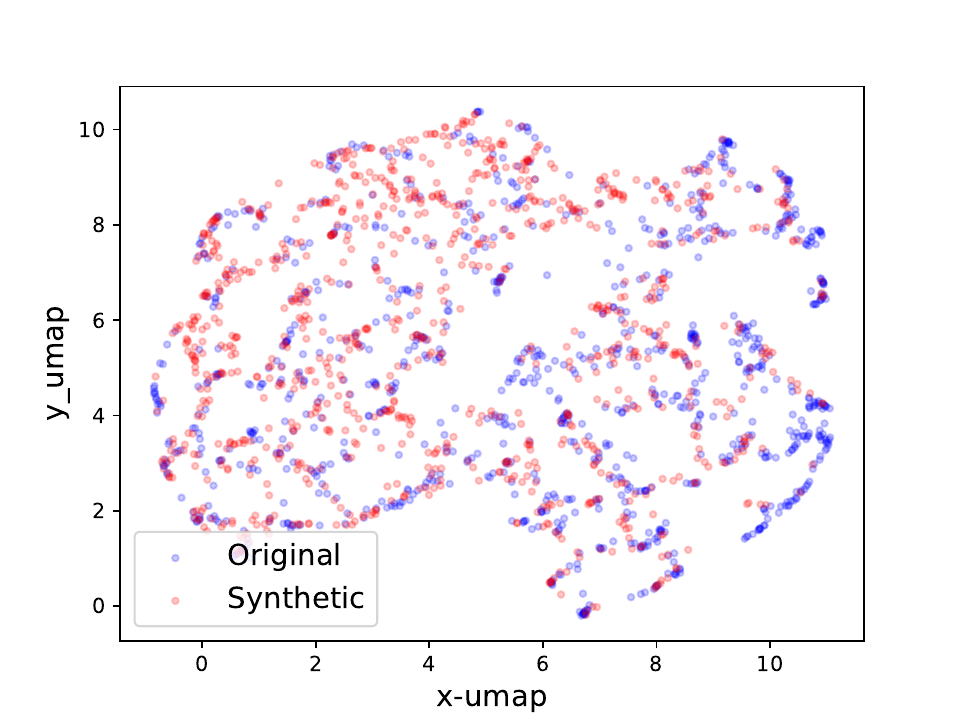}
        \caption{Running}
    \end{subfigure}
    \hfill
    \begin{subfigure}[b]{0.22\textwidth}
        \includegraphics[width=\textwidth]{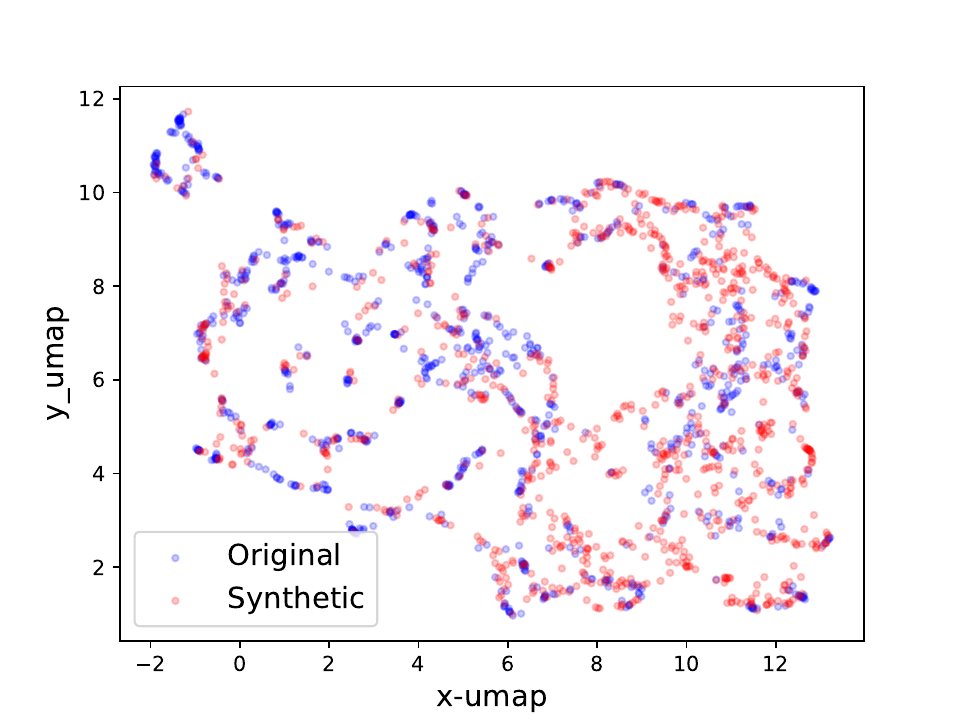}
        \caption{Going Down Stairs}
    \end{subfigure}
    \hfill
    \begin{subfigure}[b]{0.22\textwidth}
        \includegraphics[width=\textwidth]{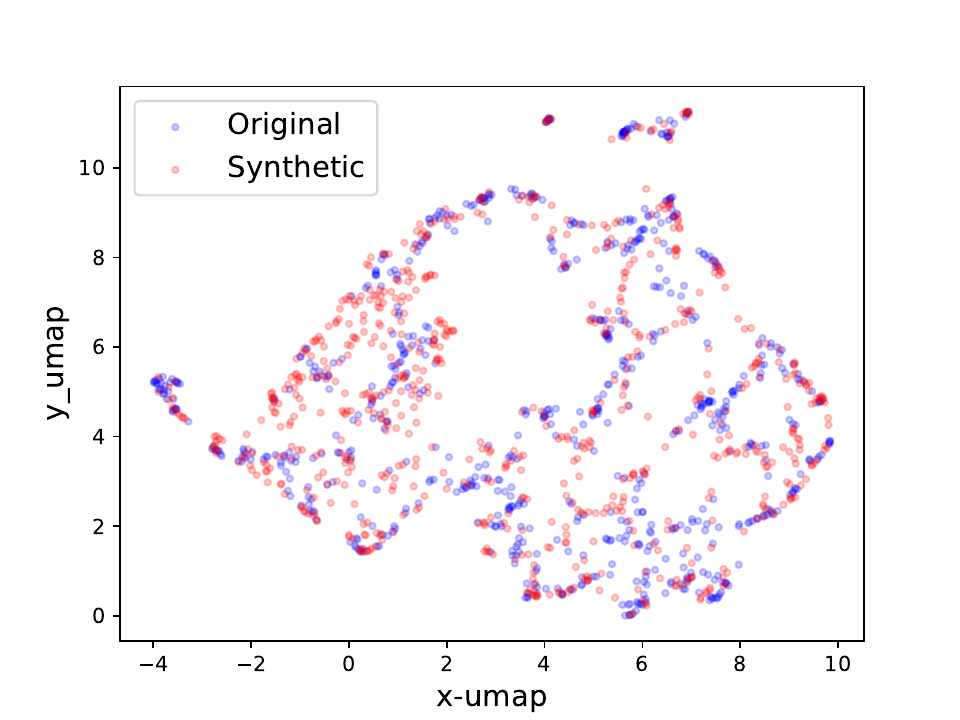}
        \caption{Going Up Stairs}
    \end{subfigure}
    \caption{Real and synthetic signal UMAP projection on selected classes of UniMiB dataset}
    \label{fig:UniMIB_umaps}
\end{figure}

\begin{figure}[htbp]
    \centering
    \begin{subfigure}[b]{0.22\textwidth}
        \includegraphics[width=\textwidth]{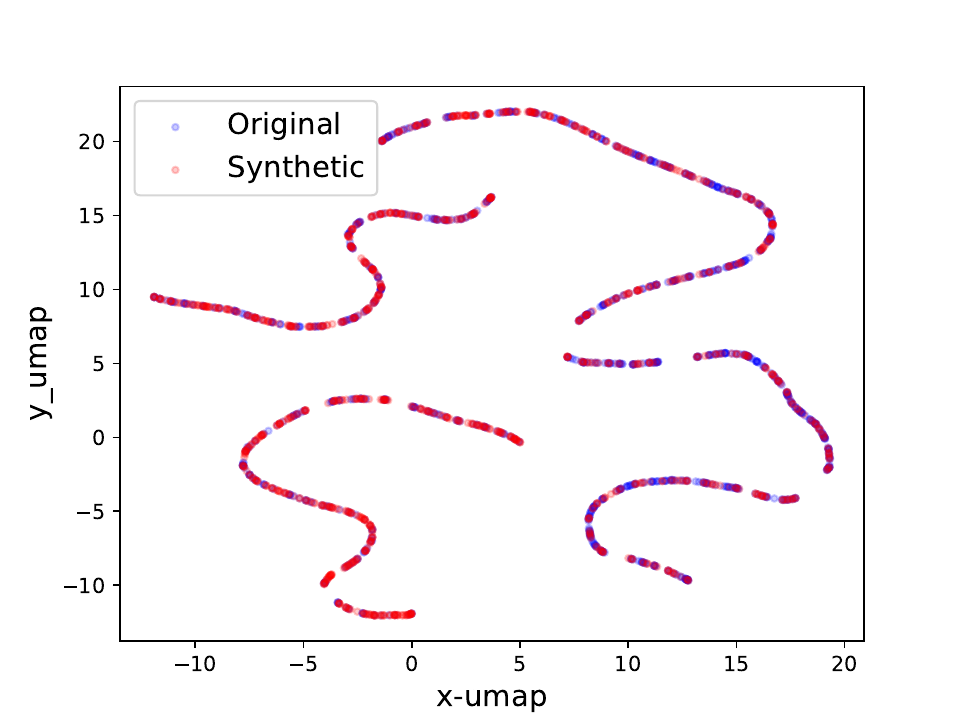}
        \caption{Class 0}
    \end{subfigure}
    \hfill
    \begin{subfigure}[b]{0.22\textwidth}
        \includegraphics[width=\textwidth]{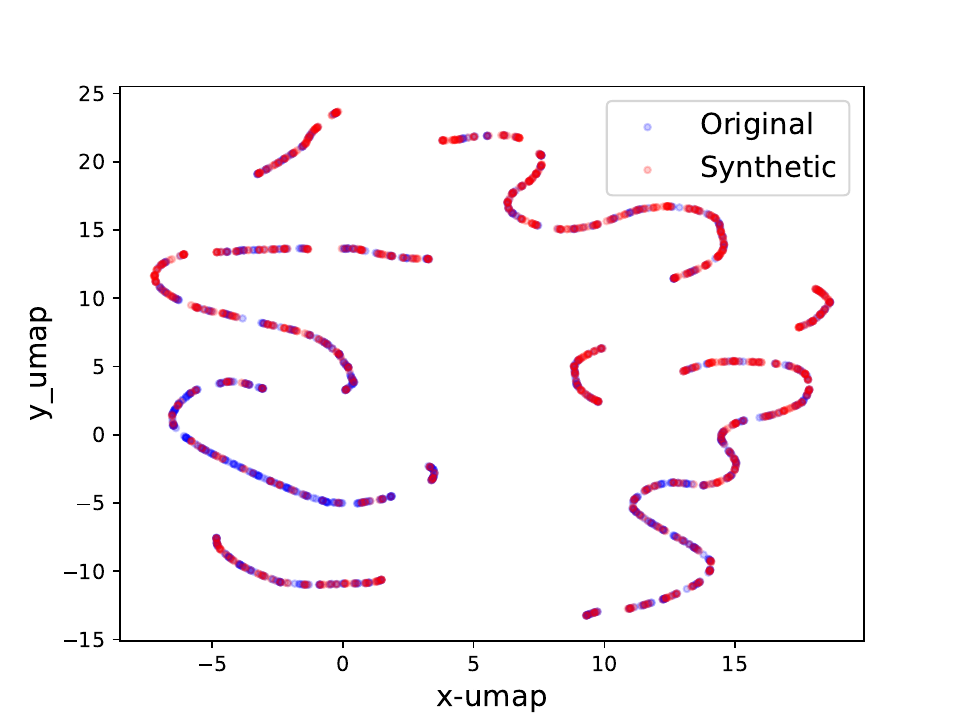}
        \caption{Class 1}
    \end{subfigure}
    \hfill
    \begin{subfigure}[b]{0.22\textwidth}
        \includegraphics[width=\textwidth]{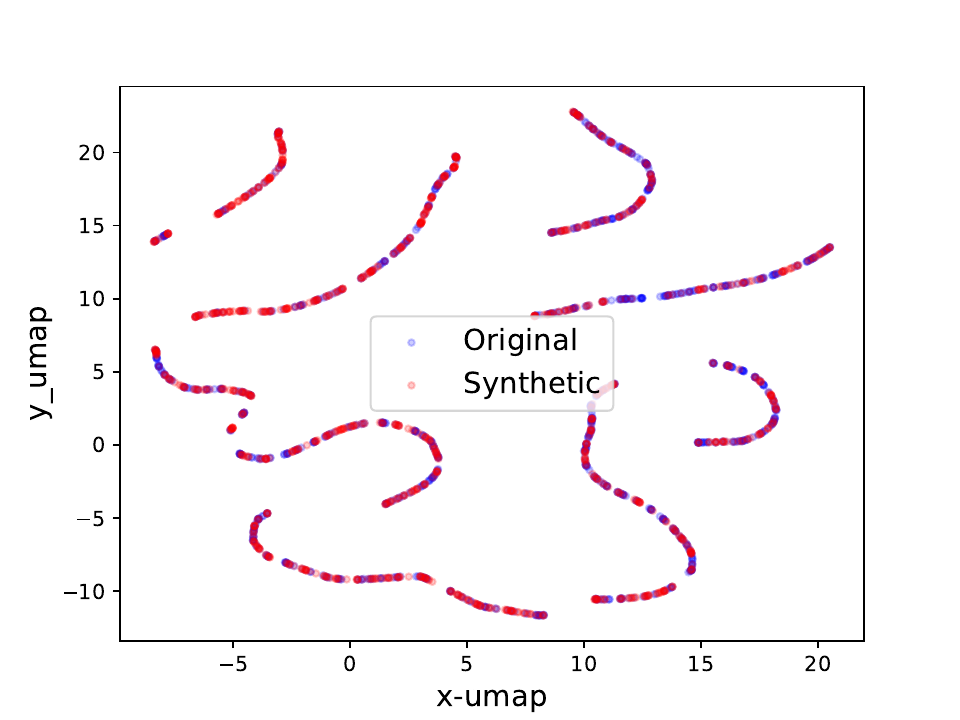}
        \caption{Class 2}
    \end{subfigure}
    \hfill
    \begin{subfigure}[b]{0.22\textwidth}
        \includegraphics[width=\textwidth]{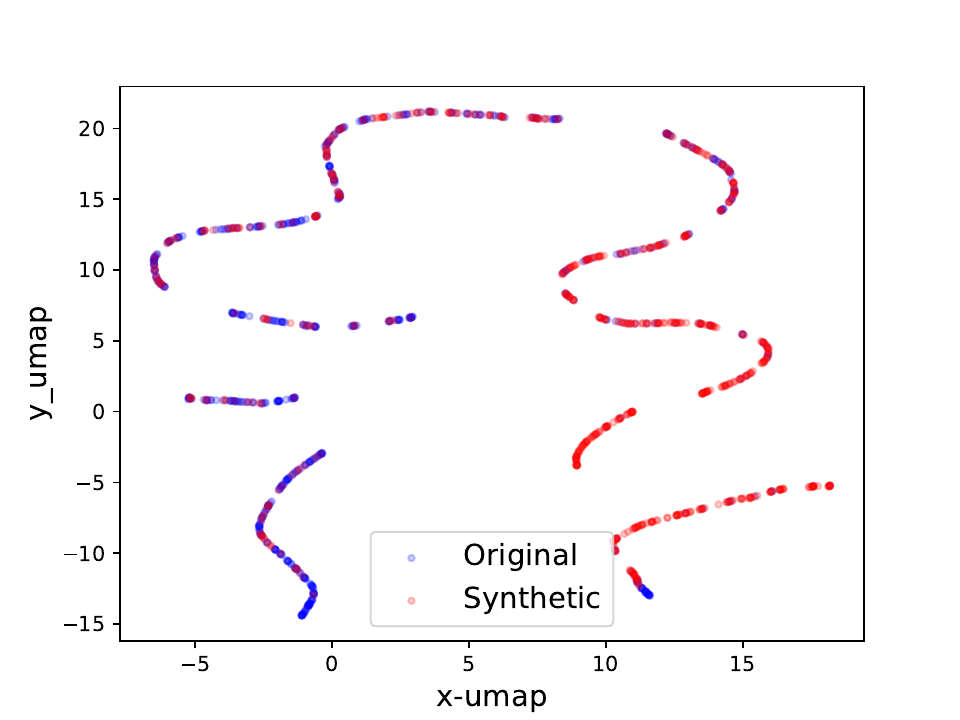}
        \caption{Class 3}
    \end{subfigure}
    \caption{Real and synthetic signal UMAP projection on selected classes of MITBIH ECG dataset}
    \label{fig:MITBIH_umaps}
\end{figure}

\section{More visualizations about signal conditional generation}
\subsection{Signal Denoising}
We selected three types of noise that are frequently involved in time-series signal collections. They are: 
\begin{itemize}
    \item \textbf{Thermal noise}, also known as white noise, is a type of random electrical noise that occurs in electronic circuits and arises from the thermal agitation of electrons, which results in a fluctuation of the voltage or current that is independent of the signal being measured.
    \item \textbf{Electrode contact noise}, also known as low-frequency drift, is a type of noise that arises in electronic measurements due to changes in the electrical characteristics of the contact between the electrode and the surface being measured, which can cause fluctuations in the baseline signal over time.
    \item \textbf{Motion artifacts}, also known as random spikes, are unwanted signals that can occur in physiological or biological measurements due to movement or other physical disturbances, which can cause sudden, brief spikes in the recorded signal that are not related to the underlying biological activity being measured.
\end{itemize}

\begin{figure}[htbp]
    \centering
    \begin{subfigure}[b]{0.8\textwidth}
        \includegraphics[width=\textwidth]{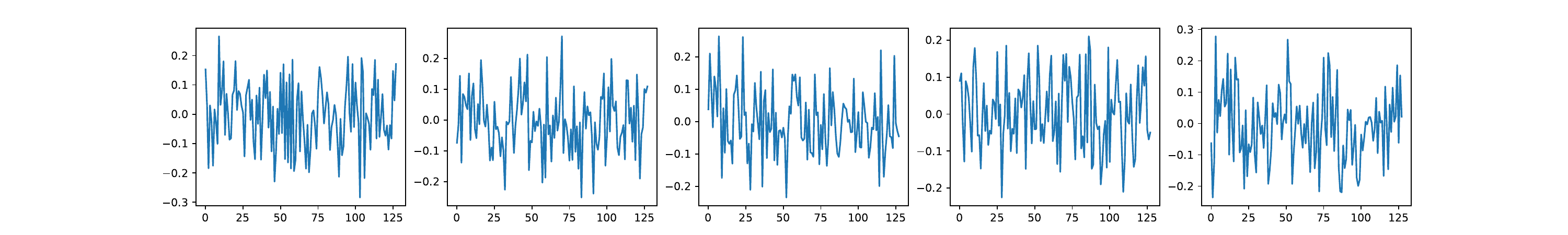}
        % \caption{Thermal noise}
        \label{fig:Thermal_noise}
    \end{subfigure}
    \begin{subfigure}[b]{0.8\textwidth}
        \includegraphics[width=\textwidth]{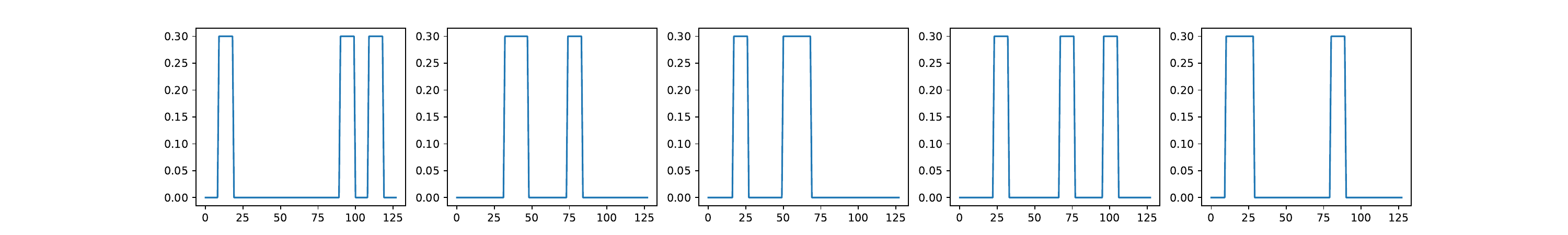}
        % \caption{Motion artifacts noise}
        \label{fig:Motion_artifacts}
    \end{subfigure}
    \begin{subfigure}[b]{0.8\textwidth}
        \includegraphics[width=\textwidth]{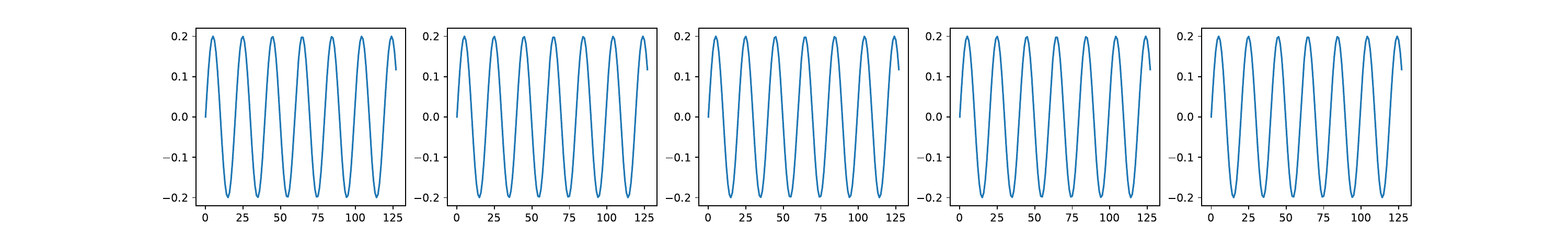}
        % \caption{Electrode noise}
        \label{fig:Electrode_noise}
    \end{subfigure}
    \caption{Three type of noises involved in biomedical signals. First row: Thermal noise. Second row: Motion artifacts noise Third row: Electrode noise }
    \label{fig:signal_noise}
\end{figure}

Fig.\ref{fig:signal_noise} shows some random noise examples. We intentionally distort the signals with such noise to make them as signal condition input to the diffusion models. The Fig.~\ref{fig:signal_denoising} shows how BioDiffusion can help remove signal artifacts. The top row shows some real signals from the MITBIH dataset. The middle row shows the same signals with added artificial noise. They are the input to the diffusion model. The bottom row shows the generated synthetic signals, which ideally should be as much as the top row signals. Please note that these signals are from the MITBIH testing set, which are unseen when training the signal conditional diffusion model. 

\begin{figure}[htbp]
    \centering
    \includegraphics[width=\textwidth]{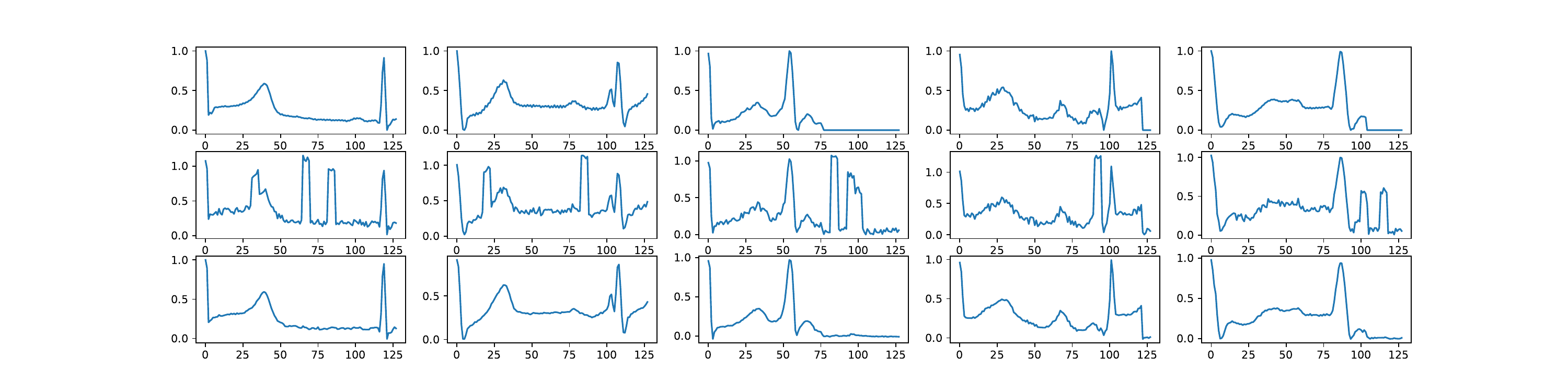}
    \caption{Example signal denoising results. First row: real signals. Second row: signals with noise. Third row: denoised signals using BioDiffusion.}
    \label{fig:signal_denoising}
\end{figure}

\subsection{Signal Imputation}
Signal imputation is another task that BioDiffusion can handle. Oftentimes, the collected signals may contain some missing values. We can use BioDiffusion to fill in those blanks. Fig.~\ref{fig:signal_imputation} shows a few examples of signal imputation. The first row shows the original signals. The second row shows the same signals with some randomly missing values (values set to zero). We use them as signal conditions input to the diffusion model. The third row shows the reconstructed signals. We can see that the synthetic signals fill in the blanks and are very similar to the original signals. 

\begin{figure}[htbp]
    \centering
    \includegraphics[width=\textwidth]{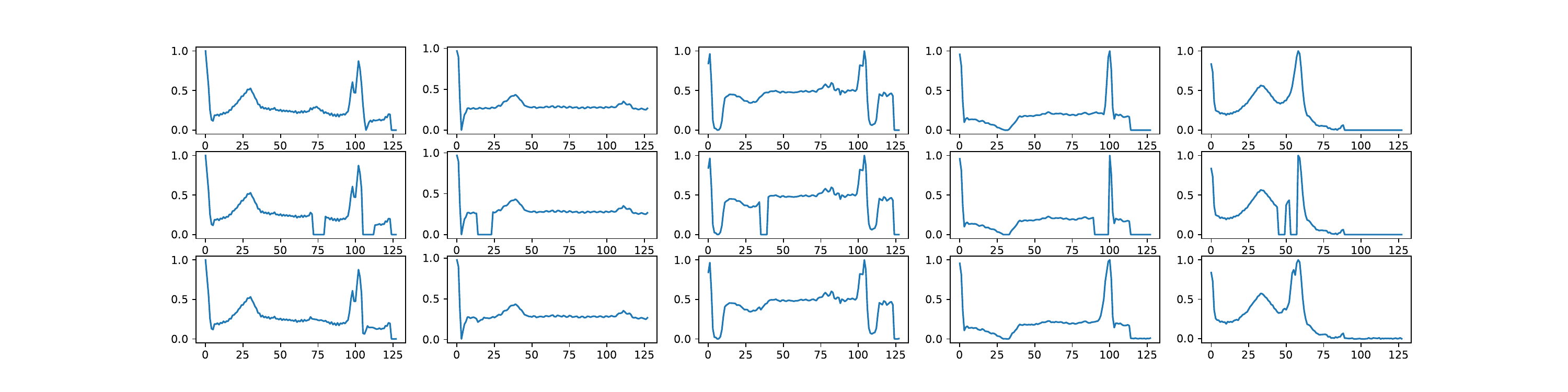}
    \caption{Example signal imputation results. First row: real signals. Second row: signals with random blanks. Third row: imputed signals using BioDiffusion.}
    \label{fig:signal_imputation}
\end{figure}

\subsection{Signal Super-resolution}

\begin{figure}[htbp]
    \centering
    \includegraphics[width=\textwidth]{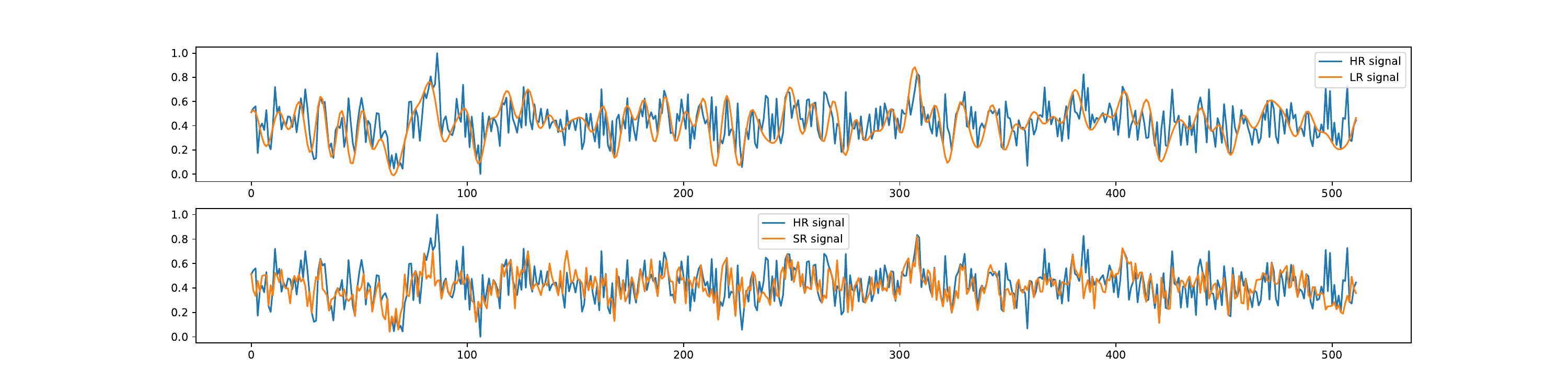}
    \caption{This figure provides an example of signal super-resolution results, where the blue lines represent one of the original signals with 512 timesteps, the upper graph orange line shows the signal downsampled to 1/4 and then upsampled to 512 timesteps using the Scikit-Learn `resample()' method, and the bottom graph orange line shows the super-resolution signal generated by the diffusion model using the downsampled signal as conditional input.}
    \label{fig:signal_sr_lr}
\end{figure}

Biomedical signals of identical types can possess distinct sampling rates due to the usage of different equipment for collection. This necessitates the application of signal downsampling or upsampling techniques to match the sampling rates when these signals are used concurrently. However, conventional upsampling methods like Hamming windows, linear/cubic interpolation, and zero-padding followed by low-pass filtering, may fall short in capturing intricate relationships among signal features. This shortcoming restricts their capacity to generate high-quality, realistic upsampled signals. A potential solution to this limitation can be found in deep learning-based super-resolution techniques. Our signal conditional diffusion model, designed for signal upsampling, has been trained to create high-resolution signals that closely resemble their original counterparts. This is illustrated in Fig~\ref{fig:signal_sr_lr}, where the model-generated signal exhibits features more akin to the original signal than the downsampled version.

\subsection{Individual signal generation}
%Lacked data samples from individual subject are another problem that impede some machine learning applications on biomedical signals. We can relieve this problem by using signal conditional diffusion models. At first, we train a diffusion model from a type of signals from many subjects. And then use a few single subject's signals as conditional inputs to let the diffusion model generate arbitrary numbers of synthetic signals that contain the unique data patterns of this subject. Therefore, we can use synthetic signals to expend the dataset size of an individual subject and develop machine learning applications specific for a single subject. 

One of the challenges that hinders certain machine learning applications on biomedical signals is the insufficient data samples from each individual subject. To address this issue, signal conditional diffusion models can be utilized. Initially, a diffusion model is trained on a specific type of signals from numerous subjects. Afterwards, a small number of signals from a single subject are utilized as conditional inputs to enable the diffusion model to generate a multitude of synthetic signals that incorporate the distinctive data patterns of that subject. As a result, synthetic signals can be employed to expand the dataset size of an individual subject and facilitate the development of machine learning applications tailored to that particular subject.

\begin{figure}[htbp]
    \centering
    \begin{subfigure}[b]{0.45\textwidth}
        \includegraphics[width=\textwidth]{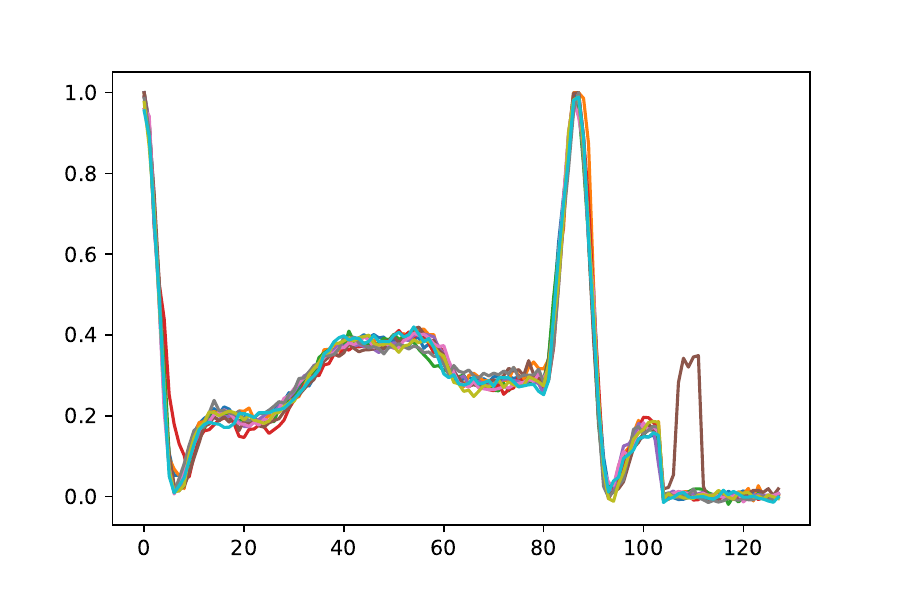}
    \end{subfigure}
    \begin{subfigure}[b]{0.45\textwidth}
        \includegraphics[width=\textwidth]{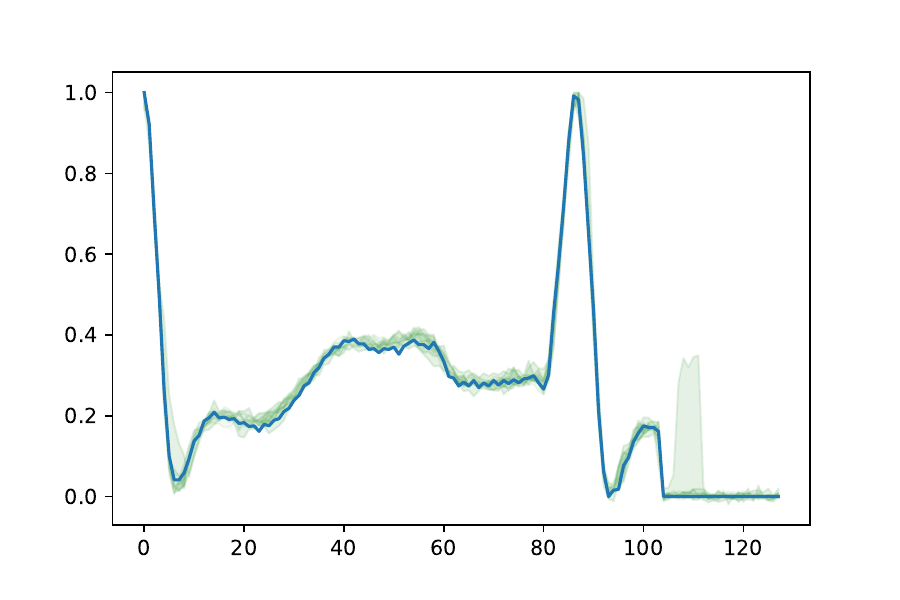}
    \end{subfigure}
    \caption{Generated synthetic heartbeat signals from a single real signal. The left graph shows 10 synthetic signals. In the right graph, the blue line represents the original signal, and the green area shows the value range of the 10 synthetic signals.}
    \label{fig:single_generation}
\end{figure}

\end{document}